\documentclass[aps, prx, twocolumn, superscriptaddress]{revtex4-1}
\usepackage{bm, amsmath, amsfonts, amssymb, braket}
\usepackage[dvipdfmx]{graphicx}
\usepackage{float}
\usepackage{color}

\usepackage{comment}
\usepackage{xspace}
\usepackage{stmaryrd}
\usepackage{ulem}

\newcommand{\ii}{\text{i}}
\newcommand{\hop}{J}

\begin{document}

\title{Parity-time-symmetric topological superconductor}

\author{Kohei Kawabata}
\email{kawabata@cat.phys.s.u-tokyo.ac.jp}
\affiliation{Department of Physics, University of Tokyo, 7-3-1 Hongo, Bunkyo-ku, Tokyo
113-0033, Japan}

\author{Yuto Ashida}
\affiliation{Department of Physics, University of Tokyo, 7-3-1 Hongo, Bunkyo-ku, Tokyo
113-0033, Japan}

\author{Hosho Katsura}
\affiliation{Department of Physics, University of Tokyo, 7-3-1 Hongo, Bunkyo-ku, Tokyo
113-0033, Japan}

\author{Masahito Ueda}
\affiliation{Department of Physics, University of Tokyo, 7-3-1 Hongo, Bunkyo-ku, Tokyo
113-0033, Japan}
\affiliation{RIKEN Center for Emergent Matter Science (CEMS), Wako, Saitama 351-0198, Japan}

\date{\today}

\begin{abstract}
We investigate a topological superconducting wire with balanced gain and loss that is effectively described by the non-Hermitian Kitaev/Majorana chain with parity-time symmetry. This system is shown to possess two distinct types of unconventional edge modes, those with complex energies and nonorthogonal Majorana zero modes. The latter edge modes cause nonlocal particle transport with currents that are localized at the edges and absent in the bulk. This anomalous particle transport results from the interplay between parity-time symmetry (non-Hermiticity) and topological superconductivity. 
\end{abstract}

\maketitle

\section{Introduction}
	\label{sec: introduction}

The past decade has witnessed a plethora of nonequilibrium phenomena with balanced gain and loss, which are described by non-Hermitian Hamiltonians with parity-time (PT) symmetry~\cite{Konotop-review, El-Ganainy-review}. PT-symmetric systems have two distinct phases, the unbroken phase with entirely real spectra and the broken phase with some eigenenergies forming complex conjugate pairs~\cite{Bender-98}. Between the two phases occurs unusual spontaneous symmetry breaking accompanied by an exceptional point, at which some eigenstates coalesce with strong nonorthogonality~\cite{Berry-04, Heiss-12}. The PT-symmetry breaking found numerous applications unique to nonconservative systems~\cite{Makris-08, Ruter-10, Eichelkraut-13, Lin-11, Regensburger-12, Peng-14, Feng-14, Hodaei-14, Gao-15, Peng-16, Li-16, Zhen-15, Hodaei-17, Chen-17}, such as unidirectional invisibility~\cite{Lin-11, Regensburger-12, Peng-14}, laser-mode selectivity~\cite{Feng-14, Hodaei-14}, and enhanced sensitivity~\cite{Hodaei-17, Chen-17}. Moreover, distinctive aspects of PT-symmetric open quantum systems were revealed~\cite{Bender-07, Gunther-08, Graefe-08, Brody-12, Chtchelkatchev-12, Tripathi-16, Tang-16, Yin-17, Lee-14, Lee-14x, Ashida-16, Kawabata-17pt, Ghatak-18, Ashida-18, Nori-18}, including speed limits~\cite{Bender-07}, entanglement~\cite{Lee-14}, and information~\cite{Kawabata-17pt}.

Recently, there has been remarkable progress in topological characterization of non-Hermitian systems~\cite{Rudner-09, Zeuner-15, Hu-11, Esaki-11, Schomerus-13, Poli-15, Weimann-17, Zhao-18, Parto-17, St-Jean-17, Obuse-16, Malzard-15, SanJose-16, Lee-16, Leykam-17, Xu-17, Gong-17, Saleur-17, LF-17a, LF-17b, LF-17c, Segev-18, Ni-18, Gong-18, KK-18} beyond the existing framework of closed systems~\cite{Kane-review}. While the hallmark of topological phases is the emergence of localized states at the boundaries as a result of topologically nontrivial bulk properties, non-Hermiticity makes the boundary states amplified (lasing)~\cite{Schomerus-13, Poli-15, Weimann-17, Zhao-18, Parto-17, St-Jean-17, Segev-18, Obuse-16} and anomalous~\cite{Lee-16}. Furthermore, emergent Majorana fermions at the edges, which are reminiscent of the conventional topological superconducting wires, were shown to persist even in the presence of gain and loss~\cite{Wang-15, Yuce-16, Menke-17, Wunner-17, Song-17}. However, little has been known about unique non-Hermitian features of PT-symmetric topological superconductors that have no Hermitian counterparts. 

This work explores nonequilibrium topological phenomena induced by the interplay between PT symmetry and topological superconductivity. We consider a topological superconducting wire with balanced gain and loss and reveal that non-Hermiticity makes its Majorana edge modes nonorthogonal, which causes nonlocal particle transport with anomalous currents that are present only at the edges. Moreover, we find unconventional edge modes with complex eigenenergies caused by PT-symmetry breaking. These complex edge modes are induced by the localized structure of gain and loss and thus essentially different from the Majorana edge modes, which originate from the nontrivial topology. We demonstrate that the complex edge modes can be identified as the additional unpaired Majorana fermions at the PT-transition point by explicitly transforming the original non-Hermitian Hamiltonian in the PT-unbroken phase to the Hermitian Hamiltonian with the same real spectrum. While most research on PT-symmetric systems has mainly focused on one-body systems, the transformation presented here is inherent in many-particle systems and thus provides a fresh view on PT-symmetry breaking in many-particle systems. 

This paper is organized as follows. In Sec.~\ref{sec: model}, we introduce the model and describe its symmetry. In Sec.~\ref{sec: edge modes}, we investigate two distinct types of edge modes, the Majorana zero edge modes and complex edge modes. We also numerically demonstrate their robustness against disorder. In Sec.~\ref{sec: pseudo-Hermiticity}, we explicitly construct the transformation from the original non-Hermitian Hamiltonian in the PT-unbroken phase to the Hermitian Hamiltonian with the same real spectrum. In Sec.~\ref{sec: edge current}, we consider the particle currents through fermionic systems induced by non-Hermiticity. After clarifying that such particle currents are generated by the nonorthogonality of quasiparticles with a simple PT-symmetric fermionic system with two sites, we demonstrate the emergence of a nonlocal current localized at the edges in the PT-symmetric topological superconductor. In Sec.~\ref{sec: conclusion}, we conclude this paper with discussions on a possible experimental setup and some outlooks. In Appendix~\ref{appendix: symmetry}, we summarize fundamental discrete symmetries and their constraints for non-Hermitian Hamiltonians. In Appendix~\ref{appendix: finite-size}, we analytically evaluate finite-size modifications of the edge modes. In Appendix~\ref{appendix: pseudo-Hermiticity}, we describe detailed calculations on pseudo-Hermiticity of the PT-symmetric topological superconductor. In Appendix~\ref{appendix: fermion numerics}, we provide generic numerics for non-Hermitian free fermions. In Appendix~\ref{appendix: two sites}, we give detailed calculations on the PT-symmetric fermionic system with two sites.

\newpage
\section{Model and symmetry}
	\label{sec: model}

We study a one-dimensional spinless $p$-wave superconductor with gain at one edge and loss at the other. The Hamiltonian reads 
\begin{eqnarray}
\hat{H}_{\rm PT}
&=& \sum_{j=1}^{L-1} \left( -\hop\,\hat{c}_{j}^{\dag} \hat{c}_{j+1} + \ii \Delta\,\hat{c}_{j} \hat{c}_{j+1} + {\rm H.c.} \right) \nonumber \\
&~&~ - \mu \sum_{j=1}^{L} \left( \hat{c}_{j}^{\dag} \hat{c}_{j} - \frac{1}{2} \right) - \ii \gamma \left( \hat{c}_{1}^{\dag} \hat{c}_{1} - \hat{c}_{L}^{\dag} \hat{c}_{L} \right),
	\label{Hamiltonian}
\end{eqnarray}
where $\hat{c}_{j}$ ($\hat{c}_{j}^{\dag}$) annihilates (creates) a fermion on site $j$, and $\hop,\,\Delta,\,\mu,\,\gamma \in \mathbb{R}$ respectively denote the hopping amplitude, the $p$\,-wave pairing gap, the chemical potential, and the balanced gain and loss. We assume $\hop,\,\Delta,\,\gamma \geq 0$ for simplicity. The Hermitian part $\hat{H}_{0} := (\hat{H}_{\rm PT} + \hat{H}_{\rm PT}^{\dag})/2$ describes the Kitaev model for topological superconductors~\cite{Kitaev-01, Alicea-review, Sato-review}, where the Majorana edge modes emerge in the topological phase $\left| \mu/2\hop \right| < 1$. It was numerically demonstrated that they persist even in the non-Hermitian system $\hat{H}_{\rm PT}$~\cite{Menke-17, Wunner-17}. Whereas Refs.~\cite{Menke-17, Wunner-17} are concerned with the fate of the conventional properties of topological superconductors in the presence of non-Hermiticity, this work explores nonequilibrium phenomena unique to PT-symmetric topological superconductors without Hermitian counterparts.

The system has PT symmetry: 
\begin{equation}
(\hat{\mathcal{P}} \hat{\mathcal{T}})\,\hat{H}_{\rm PT}\,(\hat{\mathcal{P}} \hat{\mathcal{T}})^{-1} = \hat{H}_{\rm PT},
\end{equation} 
where parity (spatial reflection) and time reversal act as 
\begin{equation}
\hat{\cal P}\,\hat{c}_{j}\,\hat{\cal P}^{-1} = \hat{c}_{L+1-j},~~
\hat{\cal T}\,\ii\,\hat{\cal T}^{-1} = -\ii.
\end{equation} 
In addition, $\hat{H}_{\rm PT}$ has particle-hole symmetry: 
\begin{equation}
(\hat{\mathcal{P}} \hat{\mathcal{C}})\,\hat{H}_{\rm PT}\,(\hat{\mathcal{P}} \hat{\mathcal{C}})^{-1} = - \hat{H}_{\rm PT},
\end{equation} 
where charge conjugation acts as 
\begin{equation}
\hat{\cal C}\,\hat{c}_{j}\,\hat{\cal C}^{-1} = \ii \hat{c}_{j}^{\dag},~~\hat{\cal C}\,\ii\,\hat{\cal C}^{-1} = -\ii.
\end{equation} 
As a combination of the above symmetries, $\hat{H}_{\rm PT}$ also has chiral symmetry: 
\begin{equation}
\hat{\cal S}\,\hat{H}_{\rm PT}\,\hat{\cal S}^{-1} = - \hat{H}_{\rm PT},
\end{equation}
with $\hat{\cal S} := (\hat{\mathcal{P}} \hat{\mathcal{T}}) (\hat{\mathcal{P}} \hat{\mathcal{C}}) = \hat{\mathcal{T}} \hat{\mathcal{C}}$. PT symmetry guarantees the presence of a complex conjugate pair of eigenenergies $( E,\,E^{*} )$ in the spectrum~\cite{Bender-98} and particle-hole symmetry leads to $( E,\,-E^{*} )$ pairs~\cite{Malzard-15, KK-18, Pikulin-12, Ge-17, Qi-18}, which together result in the quartet structure $( E,\,E^{*},\,-E,\,-E^{*} )$ (see Appendix \ref{appendix: symmetry} for details). 

In Hermitian Hamiltonians, particle-hole and chiral symmetries impose the same constraints on the real spectrum: the spectrum should be symmetric about zero energy. In non-Hermitian Hamiltonians, by contrast, they impose the different constraints on the complex spectrum: particle-hole symmetry makes the spectrum symmetric about the imaginary axis, while chiral symmetry makes the spectrum symmetric about zero energy. This distinction originates from the fact that particle-hole symmetry is antiunitary and accompanied by complex conjugation, whereas chiral symmetry is unitary and unrelated to complex conjugation~\cite{KK-18}.

\section{Edge modes}
	\label{sec: edge modes}

\subsection{Majorana zero edge modes}

The crucial feature of topological superconductors is the emergence of Majorana fermions at their boundaries~\cite{Alicea-review, Sato-review}. Especially in one dimension, the Majorana edge modes $\hat{\Psi}_{\rm zero}$ commute with the Hamiltonian, i.e.,
\begin{equation}
[ \hat{H}_{\rm PT},\,\hat{\Psi}_{\rm zero} ] = O\,(e^{-L/\xi_{\rm zero}})
	\label{eq: Majorana - Schrodinger}
\end{equation} 
with a localization length $\xi_{\rm zero} > 0$, and have finite norms even in the thermodynamic limit, i.e.,
\begin{equation}
\lim_{L \to \infty} \hat{\Psi}_{\rm zero}^{\dag} \hat{\Psi}_{\rm zero} < \infty.
	\label{eq: Majorana - normalization}
\end{equation}
The PT-symmetric superconducting wire $\hat{H}_{\rm PT}$ also possesses the Majorana edge modes in the topological phase $\left| \mu/2\hop \right| < 1$ with $\hat{\Psi}_{\rm zero}^{\rm L\,(R)}$ localized at the left (right) edge, as described in detail below. 

The Majorana edge modes in the Kitaev chain with gain/loss at its boundaries are obtained in a manner similar to those in the Kitaev chain with twisted boundaries~\cite{Creutz-99, Viola-16, Kawabata-17k}. If we introduce Majorana operators 
\begin{equation} \begin{split}
\hat{a}_{j} &:= \hat{c}_{j} e^{\ii \pi/4} + \hat{c}_{j}^{\dag} e^{-\ii \pi/4}, \\
\hat{b}_{j} &:= (\hat{c}_{j} e^{\ii \pi/4} - \hat{c}_{j}^{\dag} e^{-\ii \pi/4})/\ii,
\end{split} \end{equation}
the Hamiltonian given by Eq.~(\ref{Hamiltonian}) can be represented as 
\begin{equation} \begin{split}
\hat{H}_{\rm PT}
= \frac{\ii}{2} \left\{ \sum_{j=1}^{L-1} \left[ \left( \hop+\Delta \right) \hat{b}_{j} \hat{a}_{j+1} -  \left( \hop-\Delta \right) \hat{a}_{j} \hat{b}_{j+1} \right] \right. \\
\left. - \mu \sum_{j=1}^{L} \hat{a}_{j} \hat{b}_{j}
- \ii \gamma \left( \hat{a}_{1} \hat{b}_{1} - \hat{a}_{L} \hat{b}_{L} \right) \right\}.
	\label{eq: Hamiltonian - Majorana}
\end{split} \end{equation}
If the Majorana edge modes are expressed as $\hat{\Psi}_{\rm zero} = \sum_{j=1}^{L} ( A_{j} \hat{a}_{j} + B_{j} \hat{b}_{j} )$ with amplitudes $A_j,B_j \in \mathbb{C}$, the Schr\"odinger equation given by Eq.~(\ref{eq: Majorana - Schrodinger}) leads to
\begin{equation} \begin{split}
\left( \hop-\Delta \right) A_{j-1} + \mu A_{j} + \left( \hop+\Delta \right) A_{j+1} &= 0, \\
\left( \hop+\Delta \right) B_{j-1} + \mu B_{j} + \left( \hop-\Delta \right) B_{j+1} &= 0 
	\label{eq: zero-bulk}
\end{split} \end{equation}
in the bulk ($j=2,3\cdots,L-1$) and 
\begin{equation} \begin{split}
\left( \mu + \ii \gamma \right) A_{1} + \left( \hop+\Delta \right) A_{2} &= 0, \\
\left( \hop-\Delta \right) A_{L-1} + \left( \mu - \ii \gamma \right) A_{L} &= 0, \\
\left( \mu + \ii \gamma \right) B_{1} + \left( \hop-\Delta \right) B_{2} &=0, \\
\left( \hop+\Delta \right) B_{L-1} + \left( \mu - \ii \gamma \right) B_{L} &= 0
	\label{eq: zero-edge}
\end{split} \end{equation}
at the boundaries. We here take the thermodynamic limit $L \to \infty$ (see Appendix \ref{appendix: finite-size} for details about the finite-size modifications). Since $A_{j}$ and $B_{j}$ are independent of each other, we consider $\hat{\Psi}_{\rm zero}$ described by $\hat{a}_{j}$, which is localized at the left edge ($\hat{\Psi}_{\rm zero}$ described by $\hat{b}_{j}$ is localized at the right edge). 

The bulk conditions given by Eq.~(\ref{eq: zero-bulk}) form second-order linear recurrence equations. Hence their general solutions can be written as $A_{j} = A_{+} \lambda_{+}^{j} + A_{-} \lambda_{-}^{j}$, where $\lambda_{\pm}$ are the solutions of the characteristic equation
\begin{equation}
\left( \hop + \Delta \right) \lambda_{\pm}^{2} 
+ \mu \lambda_{\pm}
+ \left( J - \Delta \right)
= 0,
\end{equation}
which leads to
\begin{equation}
\lambda_{\pm} = \frac{-\mu \pm \sqrt{\mu^{2} - 4 \left( \hop^{2} - \Delta^{2} \right)}}{2 \left( \hop+\Delta \right)}.
\end{equation}
Here the absolute values of $\lambda_{\pm}$ should be less than $1$ in order for $\hat{\Psi}_{\rm zero}^{\rm L}$ to satisfy the normalization condition given by Eq.~(\ref{eq: Majorana - normalization}). This requirement leads to $\left| \mu/2\hop \right| \leq 1$, which determines the topological phase. The boundary conditions given by Eq.~(\ref{eq: zero-edge}) provide constraints on $A_{\pm}$. In fact, the condition $\left( \mu + \ii \gamma \right) A_{1} + \left( \hop+\Delta \right) A_{2} = 0$ implies
\begin{equation}
\frac{A_{-}}{A_{+}} = - \frac{\hop-\Delta - \ii \gamma \lambda_{+}}{\hop-\Delta - \ii \gamma \lambda_{-}}.
\end{equation}
On the other hand, the condition $\left( \hop-\Delta \right) A_{L-1} + \left( \mu - \ii \gamma \right) A_{L} = 0$ always holds for the thermodynamic limit $L \to \infty$ in the topological phase ($\left| \lambda_{\pm} \right| \leq 1$).

In the case of $\hop=\Delta$, the Majorana edge modes are explicitly obtained as
\begin{equation} \begin{split}
\hat{\Psi}_{\rm zero}^{\rm L}
&\propto \hat{a}_{1} - \frac{\mu+\ii \gamma}{2\hop} \sum_{j=2}^{L} \left( - \frac{\mu}{2\hop} \right)^{j-2} \hat{a}_{j}, \\
\hat{\Psi}_{\rm zero}^{\rm R}
&\propto \hat{b}_{L} - \frac{\mu-\ii \gamma}{2\hop} \sum_{j=2}^{L} \left( - \frac{\mu}{2\hop} \right)^{j-2} \hat{b}_{L+1-j}.
	\label{eq: Majorana edge modes}
\end{split} \end{equation}
In stark contrast to Hermitian systems, particle-hole symmetry $\hat{\cal P} \hat{\cal C}$ alone cannot protect topological edge modes to possess zero energy, and their protection in non-Hermitian systems is due to chiral symmetry $\hat{\cal S}$. The crucial distinction is that particle-hole symmetry leads to either ${\rm Re}\,E = 0$ or $( E,\,-E^{*} )$ pairs in non-Hermitian systems~\cite{Malzard-15, KK-18, Pikulin-12, Ge-17, Qi-18}, whereas it leads to either $E = 0$ or $( E,\,-E )$ pairs in Hermitian systems. As a result, topological edge modes protected by particle-hole symmetry are not restricted to zero energy and can have pure imaginary energies~\cite{Ge-17}. On the other hand, topological edge modes protected by chiral symmetry are restricted to zero energy even in non-Hermitian systems: if a zero mode $\hat{\Psi}_{\rm zero}$ localized at one edge is perturbed to have a nonzero energy $\delta \neq 0$, there should exist the other mode $\hat{\cal S} \hat{\Psi}_{\rm zero} \hat{\cal S}^{-1}$ localized at the same edge with energy $- \delta$, which is incompatible with the assumption that the number of topologically protected edge modes is at most one per one edge in the presence of an energy gap~\cite{Ryu-02}.

Although the Majorana edge modes $\hat{\Psi}_{\rm zero}$ persist even in the presence of gain and loss, non-Hermiticity modifies their anticommutation relations. Provided $\hop=\Delta$, $\hat{\Psi}_{\rm zero}^{\rm L}$ in Eq.~(\ref{eq: Majorana edge modes}) satisfies
\begin{eqnarray} \begin{split}
\{\hat{\Psi}_{\rm zero}^{\rm L},\,(\hat{\Psi}_{\rm zero}^{\rm L})^{\dag}\} 
&= 2 \left( 1 + \frac{\gamma^{2}}{4\hop^{2}} \right), \\
\{\hat{\Psi}_{\rm zero}^{\rm L},\,\hat{\Psi}_{\rm zero}^{\rm L}\} 
&= 2 \left( 1 + \frac{\gamma \left( 2\ii \mu - \gamma \right)}{4\hop^{2}} \right),
	\label{eq: statistics}
\end{split} \end{eqnarray}
where $\hat{\Psi}_{\rm zero}^{\rm L}$ is normalized so that $\hat{\Psi}_{\rm zero}^{\rm L}$ satisfies the canonical Majorana fermion anticommutation relations in the Hermitian limit ($\gamma = 0$). These anomalous anticommutation relations make a sharp contrast with the canonical anticommutation relations for ordinary Majorana fermions. They arise from the nonorthogonality of eigenstates of non-Hermitian Hamiltonians~\cite{Brody-14} and generate an anomalous particle current unique to non-Hermitian systems, as described in Sec.~\ref{sec: edge current}.

\subsection{Complex edge modes}

\begin{figure}[t]
\centering
\includegraphics[width=86mm]{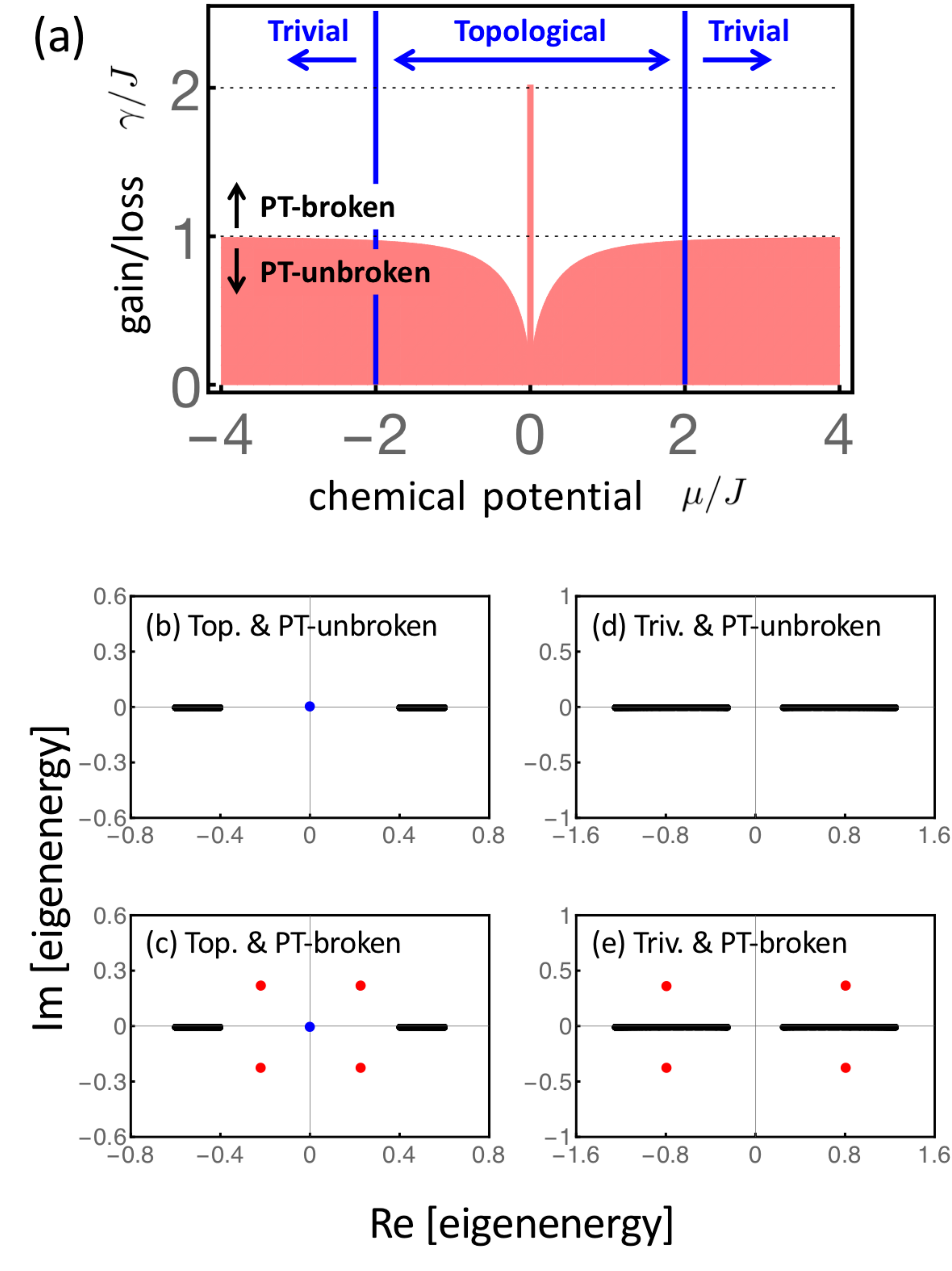} 
\caption{(a)~Phase diagram of the PT-symmetric Majorana chain with $\hop=\Delta$. The topological (trivial) phase lies in the region $\left| \mu/2\hop \right| < \left( > \right) 1$, while the PT-unbroken (-broken) phase lies in the region $\gamma < \left( > \right) \sqrt{2\,|\mu|\,( \sqrt{\hop^{2}+\mu^{2}} - |\mu| )}$. The sweet spot ($\hop=\Delta,\,\mu=0$) is singular, where the PT phase boundary lies at $\gamma = 2J$. (b-e)~Complex spectra of the PT-symmetric Majorana chain ($L = 200;\,\hop = \Delta = 0.5$). (b)~Topological and PT-unbroken phase ($\mu=0.2,\,\gamma=0.2$); Majorana zero edge modes (blue dots) emerge. (c)~Topological and PT-broken phase ($\mu=0.2,\,\gamma=1.0$); both Majorana zero edge modes (blue dots) and complex edge modes (red dots) emerge. (d)~Trivial and PT-unbroken phase ($\mu=1.5,\,\gamma=0.2$); no edge modes emerge. (e)~Trivial and PT-broken phase ($\mu=1.5,\,\gamma=1.0$); complex edge modes (red dots) emerge.}
	\label{phase diagram}
\end{figure}

Apart from the Majorana zero edge modes, there emerge additional edge modes with complex eigenenergies induced by PT-symmetry breaking. These complex edge modes satisfy 
\begin{equation}
[ \hat{H}_{\rm PT},\,\hat{\Psi}_{\rm complex} ] = E\,\hat{\Psi}_{\rm complex} + O\,( e^{-L/\xi_{\rm complex}} )
	\label{eq: complex - Schrodinger}
\end{equation}
with eigenenergy $E \in \mathbb{C}$ and  a localization length $\xi_{\rm complex} > 0$, and 
\begin{equation}
\lim_{L \to \infty} \hat{\Psi}_{\rm complex}^{\dag} \hat{\Psi}_{\rm complex} < \infty.
	\label{eq: complex - normalization}
\end{equation} 
Although determining eigenmodes is not feasible in general, focusing on the edges is sufficient to obtain $\hat{\Psi}_{\rm complex}$ since they are localized. 

We determine the complex edge modes localized at the left edge in the case of $\hop=\Delta$ for the sake of simplicity, but the generalization is straightforward. If the complex edge modes are expressed as 
\begin{equation}
\hat{\Psi}_{\rm complex}^{\rm L}
\propto \hat{a}_{1} + x \sum_{j=2}^{L} \lambda^{j-2} \hat{a}_{j} + y \sum_{j=1}^{L} \lambda^{j-1} \hat{b}_{j}, 
	\label{eq: complex edge modes}
\end{equation}
the Schr\"odinger equation given by Eq.~(\ref{eq: complex - Schrodinger}) leads to
\begin{equation}
- \ii \left( \mu + \ii \gamma \right) y = E,~~
\ii \left( 2\hop x + \mu + \ii \gamma \right) = Ey
	\label{eq: complex-edge}
\end{equation}
at the left edge, and
\begin{equation} \begin{split}
- \ii \left( 2\hop + \mu \lambda \right) \lambda^{j-2}y &= E \lambda^{j-2} x, \\
\ii \left( 2\hop + \mu \lambda^{-1} \right) \lambda^{j-1} x &= E \lambda^{j-1}y
	\label{eq: complex-bulk}
\end{split} \end{equation}
in the bulk ($j \geq 2$). We here take the semi-infinite limit $L \to \infty$ and neglect the effect of the right edge (see Appendix \ref{appendix: finite-size} for details about the finite-size modifications). If $\mu \neq 0$ is assumed, both Eq.~(\ref{eq: complex-edge}) and Eq.~(\ref{eq: complex-bulk}) imply
\begin{equation}
\lambda = \frac{2\hop\mu}{\gamma \left( 2\ii \mu - \gamma \right)},
\end{equation}
and 
\begin{equation} \begin{split}
x = \frac{2\hop \left( \mu + \ii \gamma \right)}{\gamma \left( 2\ii \mu - \gamma \right)},~~
y = \pm \left( - 1 - \frac{4\hop^{2}}{\gamma \left( 2\ii \mu - \gamma \right)} \right)^{1/2}.
\end{split} \end{equation}
To ensure the normalization condition given by Eq.~(\ref{eq: complex - normalization}), the localization length $\xi_{\rm complex} = - \left( \log \left| \lambda \right| \right)^{-1}$ should be positive. We thus need 
\begin{equation}
\gamma > \sqrt{2\,|\mu|\,( \sqrt{\hop^{2}+\mu^{2}} - |\mu| )}
\end{equation}
for the presence of the complex edge modes, which also specifies the PT-broken phase (Fig.~\ref{phase diagram}\,(a))~\cite{remark-complex-edge}. The localization length $\xi_{\rm complex}$ diverges at the PT-transition point, where the complex edge modes coalesce and form exceptional points, and gets shorter with increasing the non-Hermiticity $\gamma$. We here remark that the sweet spot ($\hop=\Delta,\,\mu=0$) is singular in the phase diagram, at which $\lambda$ is zero and the Majorana edge modes are perfectly localized \cite{Viola-16, Kawabata-17k}, and the PT phase boundary lies not at $\gamma = 0$ but at $\gamma = 2\hop$.

The complex edge modes can shift the imaginary part of many-particle eigenenergies. As a result, their presence  enhances the number of occupied fermions at the edges with time, which is a counterpart in many-particle fermionic systems to the lasing phenomena in classical~\cite{Schomerus-13, Poli-15, Weimann-17, Zhao-18, Parto-17, St-Jean-17, Segev-18} and quantum~\cite{Obuse-16} optics. We also note that the complex edge modes come in pairs $( E,\,-E )$ at one edge by chiral symmetry~\cite{Ryu-02} and in pairs $( E^{*},\,-E^{*} )$ at the other edge by PT symmetry (Fig.~\ref{phase diagram}\,(b-e)).

\subsection{Robustness against disorder}

\begin{figure}[t]
\centering
\includegraphics[width=68mm]{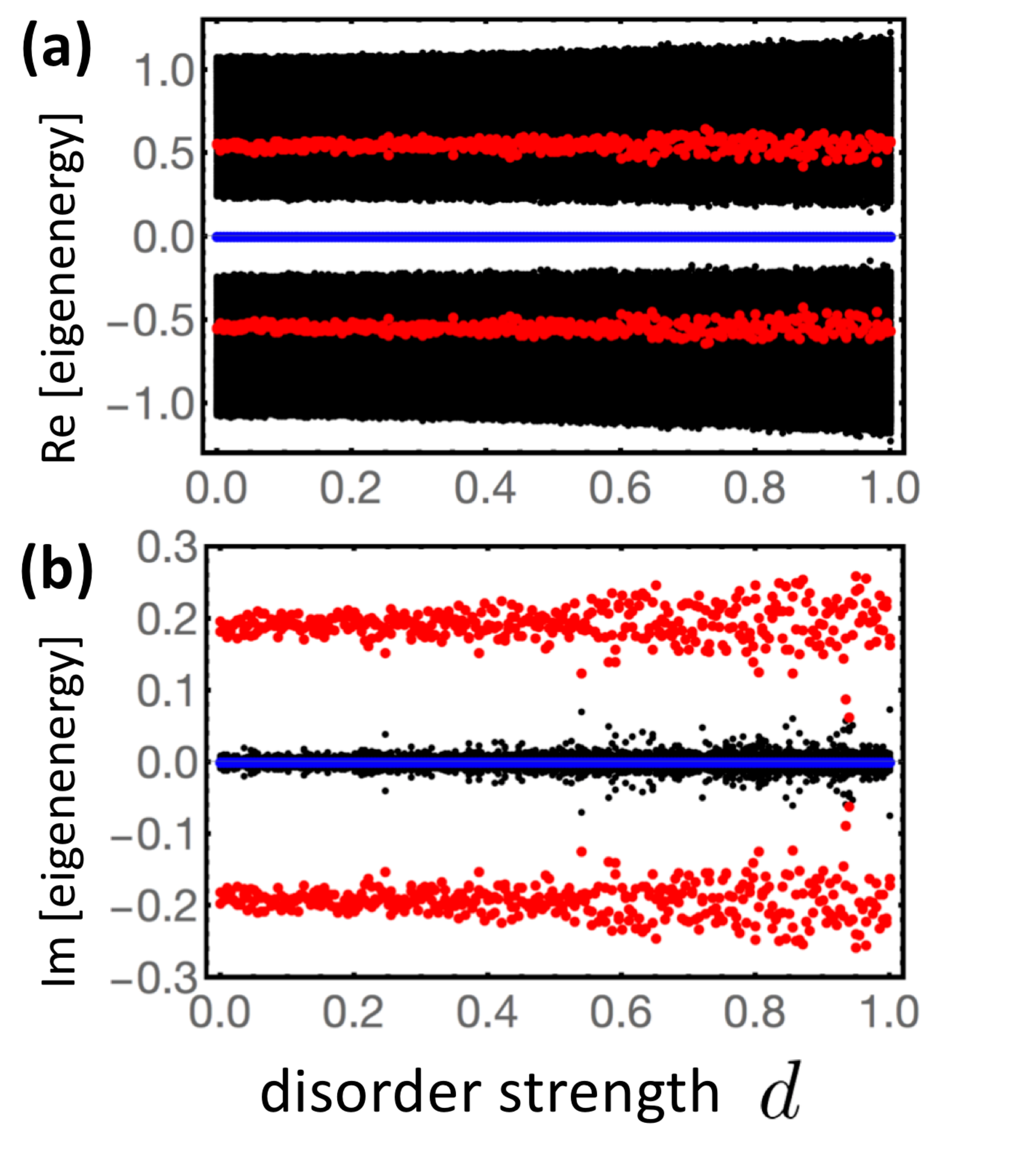} 
\caption{(a) Real and (b) imaginary parts of the single-particle spectrum as a function of the disorder strength $d$. Black, red, and blue dots represent the bulk modes, complex edge modes, and Majorana edge modes, respectively. The Majorana chain of $L=200$ sites is characterized by the parameters $\hop_{j}=0.5+0.2\epsilon_{j}$, $\Delta=1.0$, $\gamma=1.0$, and $\mu_{j} = 0.5 + d\epsilon_{j}'$, where $\hop_{j}$ is the disordered hopping amplitude between sites $j-1$ and $j$, $\mu_{j}$ is the disordered chemical potential on site $j$, and $\epsilon_{j}$ and $\epsilon_{j}'$ are uniform random variables over $\left[ -0.5,\,0.5 \right]$.}
	\label{disorder}
\end{figure}

Both Majorana zero edge modes and complex edge modes are immune to disorder. We verify this by numerically finding the single-particle spectrum for the system with disordered parameters as shown in Fig.~\ref{disorder}. Such disorder breaks PT symmetry and particle-hole symmetry, but the system nevertheless preserves chiral symmetry. Consequently, the Majorana edge modes are topologically protected with chiral symmetry against disorder (Fig.~\ref{disorder}\,(a)). On the other hand, the eigenenergies of the complex edge modes vary with disorder and hence they are not topologically protected (Fig.~\ref{disorder}\,(b)). Even so, there is a substantial imaginary gap between the complex edge modes and the bulk modes; the number of occupied fermions at the edges is enhanced with time even in the presence of disorder.

\section{Pseudo-Hermiticity}
	\label{sec: pseudo-Hermiticity}

An important consequence of unbroken PT symmetry is the existence of the Hermitian Hamiltonian $\hat{H}_{\rm H}$ that has the same real spectrum as $\hat{H}_{\rm PT}$. Such a Hermitian Hamiltonian can be constructed with a pseudo-Hermiticity operator $\hat{\eta}$~\cite{Mostafazadeh-02, Bender-02}, which is Hermitian and invertible, and satisfies 
\begin{equation}
\hat{\eta} \hat{H}_{\rm PT} = \hat{H}_{\rm PT}^{\dag} \hat{\eta}.
	\label{eq: pseudo-Hermiticity - definition}
\end{equation}
The reality of the spectrum of $\hat{H}_{\rm PT}$ is equivalent to the positivity of $\hat{\eta}$, and $\hat{H}_{\rm H}$ can be obtained with $\hat{\eta}$ as 
\begin{equation}
\hat{H}_{\rm H} = \hat{\eta}^{1/2} \hat{H}_{\rm PT} \hat{\eta}^{-1/2}.
	\label{eq: pseudo-Hermiticity - Hermitian}
\end{equation} 
In principle, $\hat{\eta}$ is determined as $\hat{\eta} = \sum_{n} \ket{\psi_{n}} \bra{\psi_{n}}$, where $\ket{\psi_{n}}$ is a right eigenstate and the summation runs over all $\ket{\psi_{n}}$. In practice, however, its analytical form is difficult to obtain for systems with multiple degrees of freedom, let alone for many-particle systems; exact forms of $\hat{\eta}$ have never been constructed for many-particle systems, although pseudo-Hermiticity is fundamentally important for PT-symmetric systems and has been studied in several respects~\cite{Mostafazadeh-02, Bender-02, Mostafazadeh-03, Bender-04, Korff-07, Fring-09, Brody-16, remark-pseudo-Hermiticity}.

Remarkably, we explicitly obtain $\hat{\eta}$ for $\hat{H}_{\rm PT}$ with odd $L$ and $\mu = 0$ in the form of the following nonlocal string operator:
\begin{eqnarray}
&\hat{\eta}&~= \left( 1 - \frac{\ii \gamma}{\hop+\Delta} \hat{a}_{1} \hat{a}_{2} \right) \left( 1 - \frac{\ii \gamma}{\hop+\Delta} \hat{b}_{2} \hat{b}_{3} \right) \cdots \nonumber \\
&~&\cdots \left( 1 - \frac{\ii \gamma}{\hop+\Delta} \hat{a}_{L-2} \hat{a}_{L-1} \right) \left( 1 - \frac{\ii \gamma}{\hop+\Delta} \hat{b}_{L-1} \hat{b}_{L} \right).~~~~~
	\label{eq: pseudo-Hermiticity}
\end{eqnarray}
It is straightforward to confirm that Eq.~(\ref{eq: pseudo-Hermiticity}) satisfies Eq.~(\ref{eq: pseudo-Hermiticity - definition}) (see Appendix~\ref{appendix: pseudo-Hermiticity} for details) and that it is positive in the PT-unbroken phase ($\gamma < \hop+\Delta$). Since $\hat{\eta}$ is composed of the products of up to $2 \left( L-1 \right)$ Majorana operators, it is nonlocal despite the locality of the non-Hermitian terms $\ii \hat{\Gamma} := ( \hat{H}_{\rm PT} - \hat{H}_{\rm PT}^{\dag} )/2 = -\ii \gamma\,( \hat{c}_{1}^{\dag} \hat{c}_{1} - \hat{c}_{L}^{\dag} \hat{c}_{L} )$. This nonlocality originates from the coexistence of a commutator and an anticommutator in the pseudo-Hermiticity algebra: $[ \hat{\eta},\,\hat{H}_{0} ] + \ii\,\{ \hat{\eta},\,\hat{\Gamma} \}=0$.

\begin{figure}[t]
\centering
\includegraphics[width=72mm]{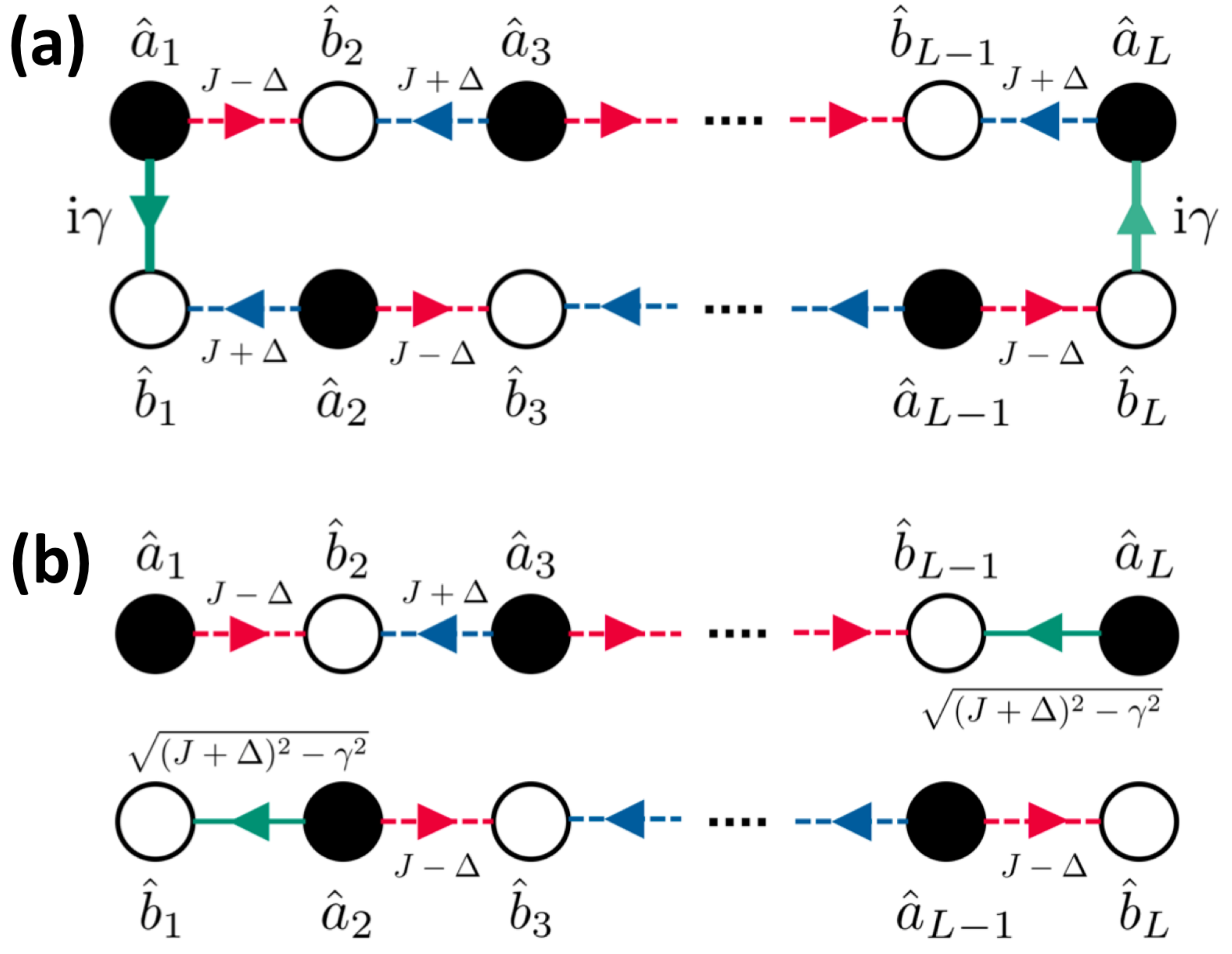} 
\caption{Schematic representations of (a) the original non-Hermitian Majorana chain $\hat{H}_{\rm PT}$ and (b) the accompanying Hermitian Majorana chain $\hat{H}_{\rm H}$ constructed with $\hat{\eta}$. Here $\hat{a}_{j}$ and $\hat{b}_{j}$ represent Majorana operators, and the lines connecting them show their couplings. As we approach the PT-transition point ($\gamma = \hop+\Delta$), the complex edge modes appear in the original representation (a), whereas the additional unpaired Majorana fermions $\hat{b}_{1}$ and $\hat{a}_{L}$ emerge in the transformed representation (b).} 
	\label{Hermitian}
\end{figure}

With the obtained $\hat{\eta}$, we have (see Appendix~\ref{appendix: pseudo-Hermiticity} for derivations)
\begin{eqnarray}
\hat{H}_{\rm H} &=& \frac{\ii}{2} \left[ \left( \hop+\Delta \right) \sum_{j=2}^{L-2} \hat{b}_{j} \hat{a}_{j+1} - \left( \hop-\Delta \right) \sum_{j=1}^{L-1} \hat{a}_{j} \hat{b}_{j+1} \right. \nonumber \\
&~&~~~~~~~\left. 
+ \sqrt{(\hop+\Delta)^{2}-\gamma^{2}}\,(\hat{b}_{1} \hat{a}_{2} + \hat{b}_{L-1} \hat{a}_{L}) \right].
	\label{eq: Hamiltonian - Hermitian}
\end{eqnarray}
The balanced gain and loss appear as the couplings $\hat{a}_{1}$-$\hat{b}_{1}$ (i.e., the coupling between $\hat{a}_{1}$ and $\hat{b}_{1}$) and $\hat{a}_{L}$-$\hat{b}_{L}$ with their amplitudes $\gamma$ in the original non-Hermitian Hamiltonian $\hat{H}_{\rm PT}$ (see Eq.~(\ref{eq: Hamiltonian - Majorana}) and Fig.~\ref{Hermitian}\,(a)), while they appear as the couplings $\hat{b}_{1}$-$\hat{a}_{2}$ and $\hat{b}_{L-1}$-$\hat{a}_{L}$ with their amplitudes $\sqrt{(\hop+\Delta)^{2}-\gamma^{2}}$ in the accompanying Hermitian Hamiltonian $\hat{H}_{\rm H}$ (see Eq.~(\ref{eq: Hamiltonian - Hermitian}) and Fig.~\ref{Hermitian}\,(b)). The Majorana edge modes protected by topology are localized near $\hat{a}_{1}$ and $\hat{b}_{L}$ in $\hat{H}_{\rm H}$. If we increase the gain/loss $\gamma$ from $0$ to $\hop+\Delta$ and approach the PT-transition point, the couplings $\hat{b}_{1}$-$\hat{a}_{2}$ and $\hat{b}_{L-1}$-$\hat{a}_{L}$ gradually decrease and vanish at the transition point; the additional unpaired Majorana fermions $\hat{b}_{1}$ and $\hat{a}_{L}$ emerge at the PT-transition point. We note that $\hat{H}_{\rm PT}$ is defective at the transition point (exceptional point) where the complex edge modes coalesce and linearly depend on each other; the additional Majorana edge modes in $\hat{H}_{\rm H}$ reflect these lost degrees of freedom.

\newpage
\section{Particle current}
	\label{sec: edge current}

It is intuitively reasonable to expect the generation of currents through the wire with gain and loss.
We hence investigate behavior of a local particle current operator
\begin{equation}
\hat{I}_{j} := -\ii \hop\,( \hat{c}_{j}^{\dag} \hat{c}_{j-1} - \hat{c}_{j-1}^{\dag} \hat{c}_{j} )
\end{equation} 
between sites $j-1$ and $j$. This quantity evaluates the difference between the hopping from site $j-1$ to site $j$ and that from site $j$ to site $j-1$, and hence can be interpreted as a particle (or electric) current in a lattice model~\cite{Furusaki-94, Scott-12, Caio-15, Ashida-18}. The current is evaluated by $I_{j} \left( t \right) := \braket{\Psi \left( t \right) | \hat{I}_{j} | \Psi \left( t \right)}$, where $\ket{\Psi \left( t \right)}$ is a many-particle wavefunction obeying the PT dynamics~\cite{Brody-12, Kawabata-17pt}
\begin{equation}
\ket{\Psi \left( t \right)} = \frac{\ket{\tilde{\Psi} \left( t \right)}}{\sqrt{\braket{\tilde{\Psi} \left( t \right) | \tilde{\Psi} \left( t \right)}}},~~
\ket{\tilde{\Psi} \left( t \right)} := e^{-\ii \hat{H}_{\rm PT} t} \ket{\Psi \left( 0 \right)}.
\end{equation} 
In the following discussion, we take an initial state as a superposition of excited states with one quasiparticle for the sake of simplicity:
\begin{equation}
\ket{\Psi \left( 0 \right)}
= \sum_{j=1}^{L} \lambda_{j} \hat{p}^{\dag}_{j} \ket{\Omega},
	\label{eq: initial state}
\end{equation}
where $\lambda_{j}$'s are nonzero coefficients, the operators $\hat{p}^{\dag}_{j}$'s create quasiparticles, and $\ket{\Omega}$ is the many-particle vacuum (ground state). See Appendix~\ref{appendix: fermion numerics} for detailed numerics of non-Hermitian free fermions.

\subsection{PT-symmetric fermionic system with two sites}

The particle current originates from the nonorthogonality~\cite{Brody-14} of quasiparticles in non-Hermitian systems. To clarify the underlying physics, we first consider a PT-symmetric fermionic system with two sites
\begin{equation}
\hat{H}_{\rm PT}^{(2)} = -\hop\,(\hat{c}_{1}^{\dag} \hat{c}_{2} + \hat{c}_{2}^{\dag} \hat{c}_{1}) - \ii \gamma\,(\hat{c}_{1}^{\dag} \hat{c}_{1} - \hat{c}_{2}^{\dag} \hat{c}_{2}),
	\label{eq: Hamiltonian - two sites}
\end{equation} 
where $J \geq 0$ denotes the hopping amplitude between the two sites and $\gamma \geq 0$ the balanced gain and loss. We diagonalize $\hat{H}_{\rm PT}^{(2)}$ to be 
\begin{equation}
\hat{H}_{\rm PT}^{(2)} = - E_{0}\,(\hat{p}_{1}^{\dag} \hat{q}_{1} - \hat{p}_{2}^{\dag} \hat{q}_{2}),
	\label{eq: Hamiltonian diagonal - two-sites}
\end{equation}
where $E_{0} := \sqrt{\hop^{2}-\gamma^{2}}$ is the single-particle eigenenergy, and the quasiparticle operators satisfy
\begin{equation}
\{ \hat{p}_{i}^{\dag},\,\hat{q}_{j} \} = \delta_{ij},~\{ \hat{p}_{i}^{\dag},\,\hat{p}_{j}^{\dag} \} = \{ \hat{q}_{i},\,\hat{q}_{j} \} = 0.
	\label{eq: anticommutation - two sites}
\end{equation}
Here the difference between the left and right eigenstates $\hat{p}_{i} \neq \hat{q}_{i}$ ($i=1,2$) is unique to non-Hermitian systems. See Appendix~\ref{appendix: two sites} for details of the following calculations.

\begin{figure}[t]
\centering
\includegraphics[width=68mm]{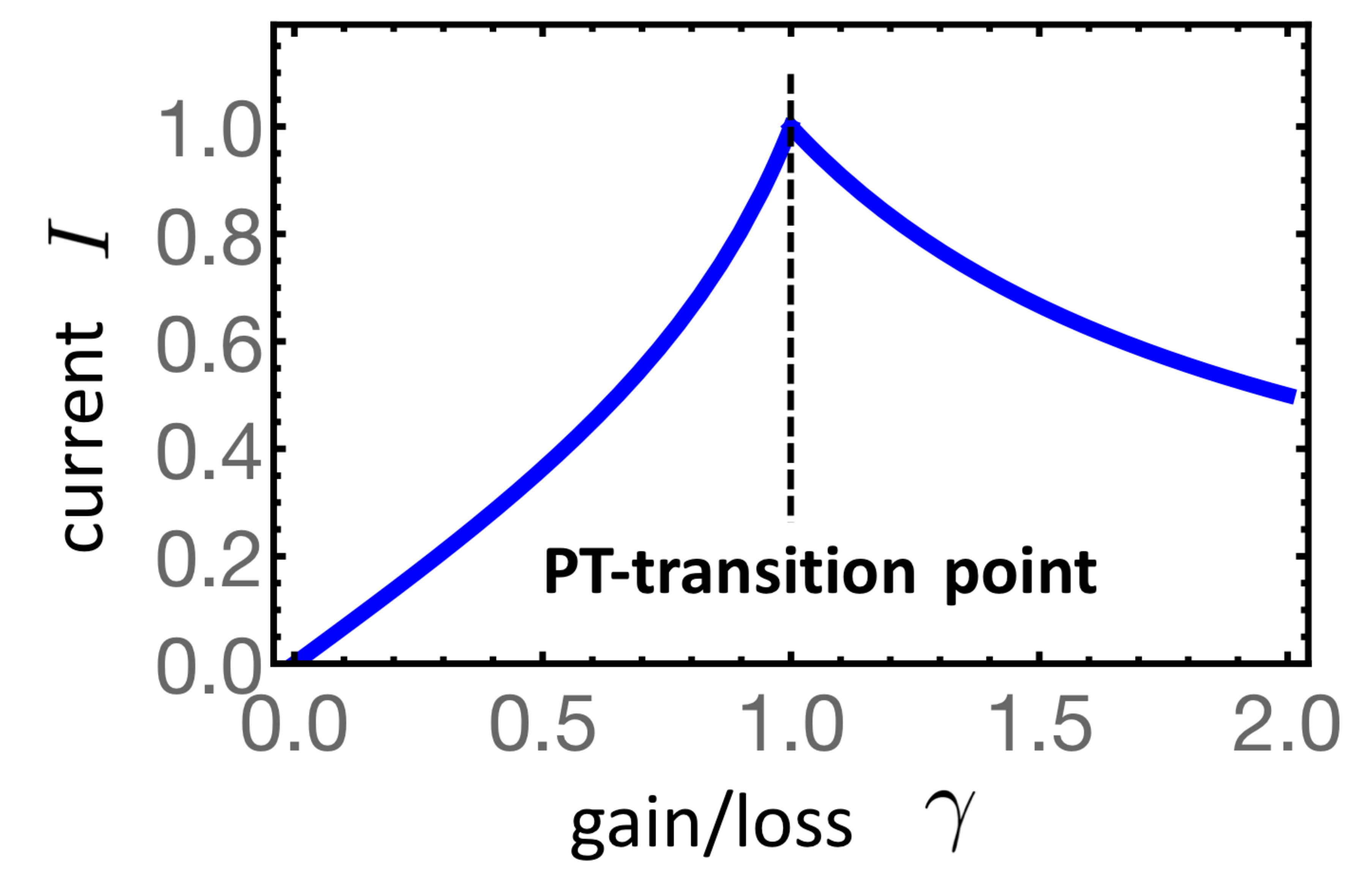} 
\caption{Current-gain/loss characteristics for the PT-symmetric fermionic system with two sites ($\hop=1.0;\,\lambda_{1} = 1/\sqrt{5},\,\lambda_{2} = 2/\sqrt{5}$). The current is the long-time average. The PT-transition point lies at $\gamma = J = 1.0$, at which the current reaches the maximum.} 
	\label{current-twolevel}
\end{figure}

In the PT-unbroken phase ($\gamma \leq J$), the quasiparticle operators satisfy
\begin{equation}
\left( \begin{array}{@{\,}c@{\,}}
      \hat{p}_{1} \\ \hat{p}_{2}
    \end{array} \right) := \left( \begin{array}{@{\,}cc@{\,}}
      1 & - \left( \ii \gamma/\hop \right) g_{r} \\ \left( \ii \gamma /\hop \right) g_{r}^{*} & 1 
    \end{array} \right) \left( \begin{array}{@{\,}c@{\,}}
      \hat{q}_{1} \\ \hat{q}_{2}
    \end{array} \right),
    	\label{eq: pq - unbroken - two-sites}
\end{equation}
where the off-diagonal term 
\begin{equation}
g_{r} := \sqrt{1- \left( \frac{\gamma}{\hop} \right)^{2}} + \frac{\ii \gamma}{\hop} 
\end{equation}
measures the degree of the nonorthogonality between $\hat{p}_{i}$ and $\hat{q}_{i}$. We take an initial state as $\ket{\Psi \left( 0 \right)} = (\lambda_{1} \hat{p}_{1}^{\dag} + \lambda_{2} \hat{p}_{2}^{\dag} ) \ket{\Omega}$, where $\ket{\Omega}$ is the vacuum for the quasiparticles (i.e., $\hat{q}_{1} \ket{\Omega} = \hat{q}_{2} \ket{\Omega} = 0$) and the coefficients satisfy $\left| \lambda_{1} \right|^{2} + \left| \lambda_{2} \right|^{2} = 1$. The particle current between the two sites is calculated to be
\begin{equation}
\frac{I \left( t \right)}{\hop}
= \frac{{\rm Im} \left[ g_{r} \right] + 2\,{\rm Im} \left[ g_{r} \lambda_{1}^{*} \lambda_{2} e^{-2\ii E_{0}t} \right]}{1 + 2 \left( \gamma/\hop \right) {\rm Im} \left[ g_{r} \lambda_{1}^{*} \lambda_{2} e^{-2\ii E_{0}t} \right]},
	\label{current - unbroken - two-sites}
\end{equation}
with $I \left( t \right) := \braket{\Psi \left( t \right) | \hat{I} | \Psi \left( t \right)}$ and $\hat{I} := -\ii \hop\,( \hat{c}_{2}^{\dag} \hat{c}_{1} - \hat{c}_{1}^{\dag} \hat{c}_{2} )$. Since the time-averaged current is approximately given by ${\rm Im} \left[ g_{r} \right] = \gamma/\hop$, the nonorthogonality characterized by $g_{r}$ produces a particle current. The current reaches the maximum at the PT-transition point ($\gamma = \hop$) because the nonorthogonality becomes maximal there (Fig.~\ref{current-twolevel}).

In the PT-broken phase ($\gamma > \hop$), on the other hand, we have
\begin{equation}
\left( \begin{array}{@{\,}c@{\,}}
      \hat{p}_{1} \\ \hat{p}_{2}
    \end{array} \right) =
\left( \begin{array}{@{\,}cc@{\,}}
      1 & g_{c} \\ g_{c} & 1
    \end{array} \right)
\left( \begin{array}{@{\,}c@{\,}}
      \hat{q}_{1} \\ \hat{q}_{2}
    \end{array} \right),
\end{equation}
where the off-diagonal term 
\begin{equation}
g_{c} := \frac{\hop}{\gamma}
\end{equation}
quantifies the degree of the nonorthogonality between $\hat{p}_{j}$ and $\hat{q}_{j}$ in the PT-broken phase. This nonorthogonality decreases with increasing gain/loss; it is maximal at the PT-transition point ($\gamma = \hop$). The particle current between the two sites is calculated to be
\begin{equation} \begin{split}
\frac{I \left( t \right)}{\hop}
\sim g_{c} \left[ 1 + 2 \left( \frac{\gamma}{\hop} - \frac{\hop}{\gamma} \right) \frac{{\rm Re} \left[ \lambda_{1}^{*} \lambda_{2} \right]}{\left| \lambda_{2} \right|^{2}} e^{-2 \sqrt{\gamma^{2}-\hop^{2}}t} \right],
	\label{current - broken - two-sites}
\end{split} \end{equation}
for $t \to \infty$. Since the current is evaluated as $I \to g_{c} J$ for a sufficiently long time, the nonorthogonality characterized by $g_{c}$ produces a particle current also in the PT-broken phase. Therefore, we conclude that the nonorthogonality between the quasiparticles induced by non-Hermiticity underlies the generation of particle (or electric) currents through fermionic systems with gain and loss.

\subsection{Nonlocal edge current}

\begin{figure}[t]
\centering
\includegraphics[width=86mm]{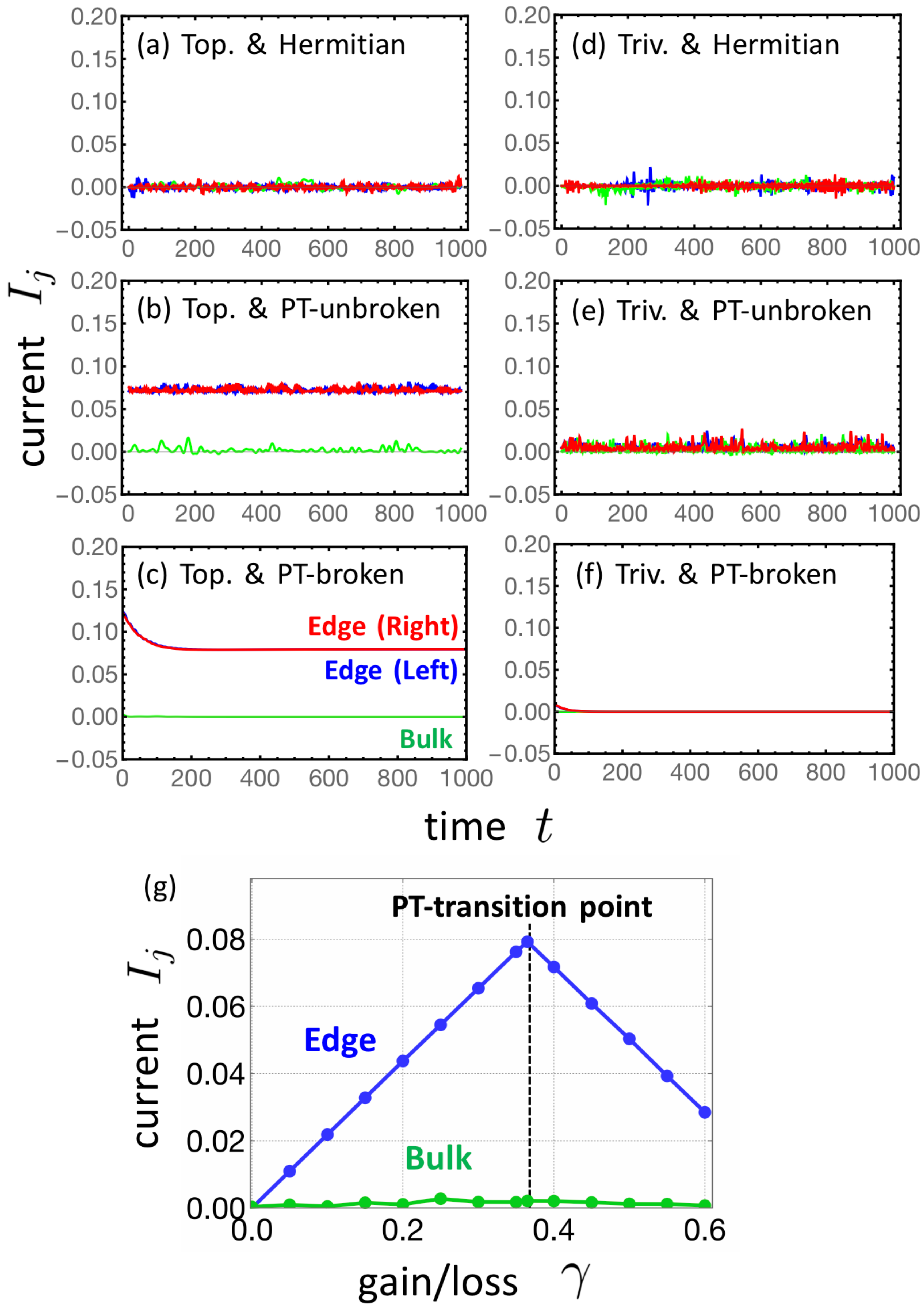} 
\caption{Particle currents through the PT-symmetric Majorana chains of $L=200$ sites with $\hop=\Delta =0.5$. The red/blue curves represent the currents at the right/left edge ($j=2$/$200$), while the green curves represent those at the center of the bulk ($j=100$). (a-c) Time evolution of the currents in the topological phase ($\mu=0.2$) for the Hermitian case ($\gamma=0.0$, (a)), the PT-unbroken phase ($\gamma=0.3$, (b)), and the PT-broken phase ($\gamma=0.4$, (c)). (d-f) Time evolution of the currents in the trivial phase ($\mu=1.5$) for the Hermitian case ($\gamma=0.0$, (d)), the PT-unbroken phase ($\gamma=0.4$, (e)), and the PT-broken phase ($\gamma=0.6$, (f)). (g) Current-gain/loss characteristics for the topological phase ($\mu=0.2$). The current is the long-time average. The PT-transition point lies at $\gamma = 0.37$, at which the edge current reaches the maximum.} 
	\label{current}
\end{figure}

\begin{figure}[t]
\centering
\includegraphics[width=86mm]{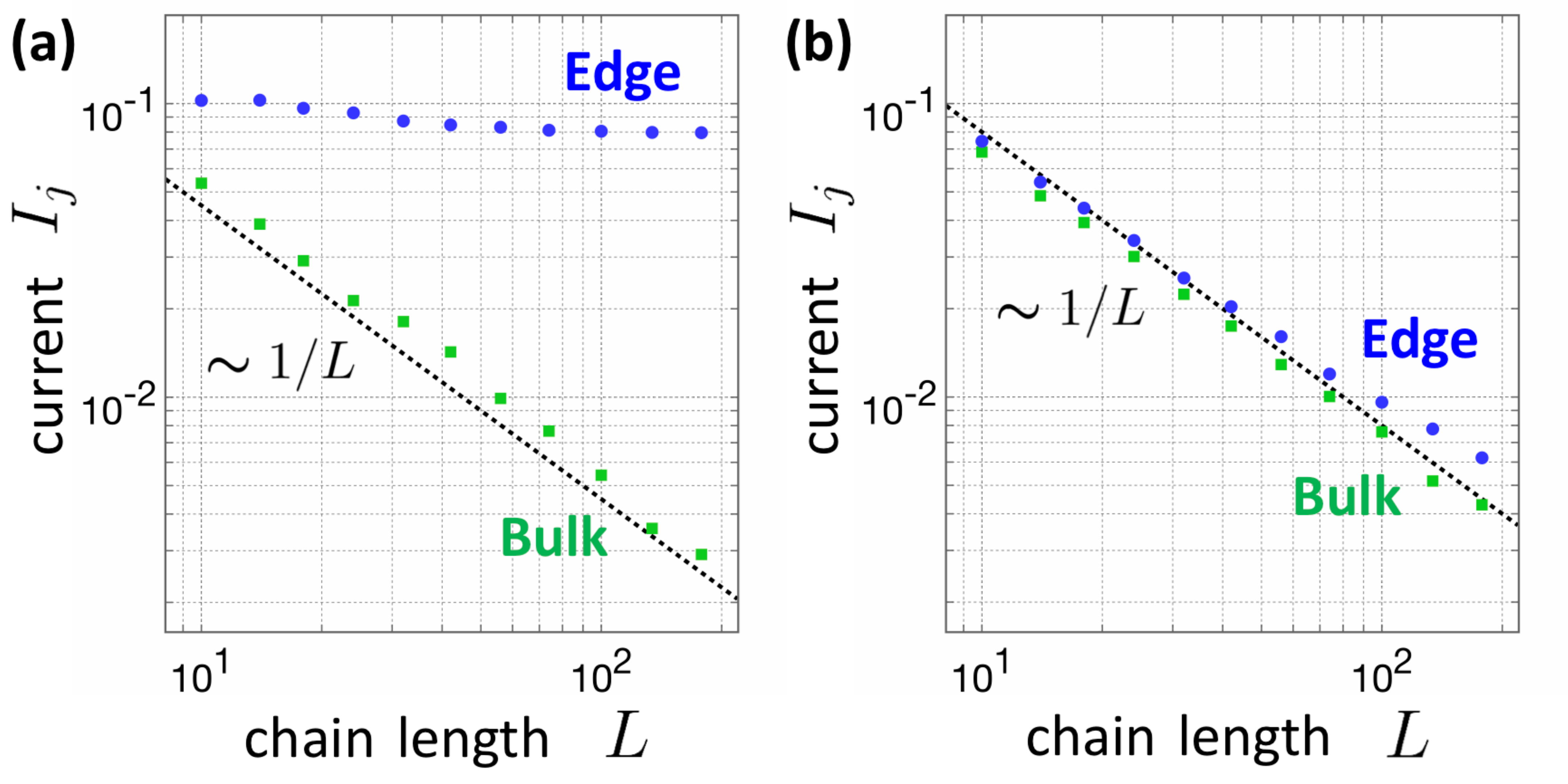} 
\caption{Finite-size scaling of the particle currents at the PT-transition point. (a) Topological phase ($\hop = \Delta = 0.5,\,\mu = 0.2,\,\gamma = 0.365$). The current in the bulk decreases according to $1/L$, whereas the current at the edge does not. (b) Trivial phase ($\hop = \Delta = 0.5,\,\mu = 1.5,\,\gamma = 0.490$). Both currents in the bulk and at the edge decrease according to $1/L$.} 
	\label{finite-size scaling}
\end{figure}

With the observation in the fermionic system with two sites, we next investigate the particle currents in the Majorana chain with balanced gain and loss. The results are summarized in Fig.~\ref{current}. In the Hermitian case (Fig.~\ref{current}\,(a,d)), there are no currents either in the bulk or at the edges, or in either topological or trivial phase. However, the situation changes in the non-Hermitian case: in the topological phase (Fig.~\ref{current}\,(b,c)), there appear nonzero currents at both edges and in both PT-unbroken and PT-broken phases, whereas no currents flow in the bulk. In the trivial phase (Fig.~\ref{current}\,(e,f)), by contrast, these currents are absent even at the edges. The edge current in the topological phase increases (decreases) with increasing gain/loss in the PT-unbroken (-broken) phase; it reaches the maximum at the PT-transition point (Fig.~\ref{current}\,(g)). Figure \ref{finite-size scaling} shows the dependence of the maximum current (current at the PT-transition point) on the chain length $L$. It is clearly seen that the current in the bulk decreases according to $1/L$, whereas the current at the edge in the topological phase does not. These results do not depend on the choice of $\lambda_{j}$'s in Eq.~(\ref{eq: initial state}).

Recalling that such particle currents are generated by the non-Hermiticity-induced nonorthogonality of quasiparticles as discussed in the last subsection, the edge current in the PT-symmetric Majorana chain is understood as follows. We first notice that non-Hermitian terms are present only at the edges in this system. In the bulk, since all the modes are delocalized, the non-Hermiticity is weak ($\sim 1/L$) and so is the nonorthogonality, which results in no currents. In fact, the currents through the bulk decrease as $1/L$ (Fig.~\ref{finite-size scaling}). At the edges, by contrast, since localized Majorana modes are present due to topology, the non-Hermiticity is not weak ($\sim 1$) and neither is the nonorthogonality, which results in the nonlocal edge current. This anomalous particle current thus originates from the combination of the nonorthogonality of quasiparticles induced by non-Hermiticity and the presence of the nonlocal Majorana edge modes induced by topology. Here the complex edge modes also contribute to the edge current; however, their contributions are much smaller than those of the Majorana edge modes since the strength of the localization of the complex edge modes is very weak near the PT-transition point, where their nonorthogonality gets maximal.

The localized particle current in one dimension is supported by the background superconducting reservoir~\cite{Fu-08, Sato-09, Lutchyn-10, Oreg-10}. Due to the pairing approximation, the conservation of particle number is locally violated; however, it is globally respected due to the nonlocal correlation between the edges produced by the Majorana edge modes. In fact, the same amount of current flows into the system at one edge and flows out from the system at the other. Therefore, our theory is consistent within the BCS approximation. It is also notable that these anomalous particle currents cannot be detected by accessing local information of the bulk alone, which is a signature of topological order.

\section{Conclusion and Discussion}
	\label{sec: conclusion}

In this work, we have studied a topological superconducting wire with balanced gain and loss. We have shown that the complex edge modes emerge at PT-symmetry breaking, while the additional Majorana edge modes emerge in the accompanying Hermitian system that has the same real spectrum as the original non-Hermitian system in the PT-unbroken phase; these properties are inherent in PT-symmetric many-particle systems. Moreover, we have unveiled the nonlocal particle current localized at the edges due to the interplay between PT symmetry and topology; this nonequilibrium topological phenomenon is unique to PT-symmetric topological systems. 

An experimental test of our prediction can be performed using fermionic ultracold atoms in a one-dimensional optical lattice, where effective $p$-wave superconductivity can be induced by creating Feshbach molecules via an optical Raman transition~\cite{Jiang-11}. Furthermore, the non-Hermiticity can be implemented by controlling and monitoring the decay of atoms~\cite{Lee-14x, Lee-14, Ashida-16}. In fact, PT-symmetry breaking has recently been observed in noninteracting ultracold fermionic $^{6}$Li atoms~\cite{Li-16}. The complex edge modes can be experimentally probed as the temporal growth of the number of occupied fermions at the edges, whereas the nonorthogonal Majorana edge modes can be observed as the nonlocal particle currents localized at the edges.

It merits further study to go beyond the BCS approximation and revisit the particle currents in a non-Hermitian interacting system that supports Majorana modes and respects particle-number conservation~\cite{Kraus-13, Ortiz-14, Lang-15, Iemini-15}. It is also of fundamental interest to investigate a thermodynamic or transport signature of edge states in non-Hermitian systems. Finally, it is worthwhile to consider topological quantum computation~\cite{Nayak-TQC} for PT-symmetric topological superconductors since non-Hermiticity effectively changes the anticommutation relations of Majorana fermions into the unconventional ones as shown in Eq.~(\ref{eq: statistics}).

\section*{Acknowledgement}

K. K. thanks Yutaka Akagi, Shunsuke Furukawa, Ryusuke Hamazaki, Sho Higashikawa, Norio Kawakami, and Masatoshi Sato for helpful discussions. This work was supported by KAKENHI Grant No.~JP18H01145, No.~JP18H04478, and No.~JP18K03445, and a Grant-in-Aid for Scientific Research on Innovative Areas ``Topological Materials Science" (KAKENHI Grant No.~JP15H05855) from the Japan Society for the Promotion of Science. K.~K. and Y.~A. were supported by the JSPS through the Program for Leading Graduate Schools (ALPS). Y.~A. acknowledges support from JSPS (KAKENHI Grant No.~JP16J03613).

\appendix

\section{Symmetries and spectra for non-Hermitian Hamiltonians}
	\label{appendix: symmetry}

We consider the constraints on a complex spectrum of a general non-Hermitian Hamiltonian $\hat{H}$ imposed by the presence of symmetries~\cite{KK-18}. If the Hamiltonian $\hat{H}$ respects PT symmetry, i.e., 
\begin{equation}
(\hat{\cal P} \hat{\cal T})\,\hat{H}\,(\hat{\cal P} \hat{\cal T})^{-1} = \hat{H},~~
(\hat{\cal P} \hat{\cal T})\,z\,(\hat{\cal P} \hat{\cal T})^{-1} = z^{*}
\end{equation}
for all $z \in \mathbb{C}$, and $\ket{\psi}$ is an eigenstate of $\hat{H}$ with eigenenergy $E \in \mathbb{C}$ (i.e., $\hat{H} \ket{\psi} = E \ket{\psi}$), we have
\begin{equation}
\hat{H}\,( \hat{\cal P} \hat{\cal T} \ket{\psi} )
= \hat{\cal P} \hat{\cal T} \hat{H} \ket{\psi}
= E^{*}\,( \hat{\cal P} \hat{\cal T} \ket{\psi} ).
\end{equation}
Hence $\hat{\cal P} \hat{\cal T} \ket{\psi}$ is an eigenstate of $\hat{H}$ with eigenenergy $E^{*}$. If $\ket{\psi}$ is also an eigenstate of $\hat{\cal P} \hat{\cal T}$, we have $E=E^{*}$, i.e., $E \in \mathbb{R}$~\cite{Bender-98}. Next, if the Hamiltonian $\hat{H}$ respects particle-hole symmetry $\hat{\cal C}$, i.e., 
\begin{equation}
\hat{\cal C} \hat{H} \hat{\cal C}^{-1} = - \hat{H},~~
\hat{\cal C}\,z\,\hat{\cal C}^{-1} = z^{*}
\end{equation}
for all $z \in \mathbb{C}$, we have
\begin{equation}
\hat{H}\,( \hat{\cal C} \ket{\psi} )
= -\hat{\cal C} \hat{H} \ket{\psi}
= - E^{*}\,( \hat{\cal C} \ket{\psi} ).
\end{equation}
Hence $\hat{\cal C} \ket{\psi}$ is an eigenstate of $\hat{H}$ with eigenenergy $-E^{*}$. If $\ket{\psi}$ is also an eigenstate of $\hat{\cal C}$, we have $E=-E^{*}$, i.e., $E \in \ii \mathbb{R}$~\cite{Malzard-15, KK-18, Pikulin-12, Ge-17, Qi-18}. We note that these constraints reduce to $E=0$ or $\left( E,\,-E \right)$ pairs in Hermitian Hamiltonians with real spectra. Finally, if the Hamiltonian $\hat{H}$ respects chiral symmetry $\hat{\cal S}$, i.e., 
\begin{equation}
\hat{\cal S} \hat{H} \hat{\cal S}^{-1} = - \hat{H},~~
\hat{\cal S}\,z\,\hat{\cal S}^{-1} = z
\end{equation} 
for all $z \in \mathbb{C}$, we have
\begin{equation}
\hat{H}\,( \hat{\cal S} \ket{\psi} )
= -\hat{\cal S} \hat{H} \ket{\psi}
= - E\,( \hat{\cal S} \ket{\psi} ).
\end{equation}
Hence $\hat{\cal S} \ket{\psi}$ is an eigenstate of $\hat{H}$ with eigenenergy $-E$. If $\ket{\psi}$ is also an eigenstate of $\hat{\cal S}$, we have $E=-E$, i.e., $E=0$. We note that these constraints remain the same for Hermitian Hamiltonians.

\onecolumngrid
\section{Finite-size modifications of edge modes}
	\label{appendix: finite-size}

In the case of $J=\Delta$, we notice that 
\begin{equation} \begin{split}
[\hat{H}_{\rm PT},\,\hat{a}_{j}]
= \begin{cases}
\ii\,( \mu + \ii \gamma)\,\hat{b}_{1} & \left( j=1 \right) \\
\ii\,( 2J \hat{b}_{j-1} + \mu \hat{b}_{j} ) & \left( 2 \leq j \leq L-1 \right) \\
\ii\,( 2J \hat{b}_{L-1} + ( \mu - \ii \gamma)\,\hat{b}_{L} ) & \left( j=L \right)
\end{cases},~~
[\hat{H}_{\rm PT},\,\hat{b}_{j}]
= \begin{cases}
-\ii\,(\,( \mu + \ii \gamma)\,\hat{a}_{1} + 2J \hat{a}_{2} ) & \left( j=1 \right) \\
-\ii\,( \mu\hat{a}_{j} + 2J \hat{a}_{j+1} ) & \left( 2 \leq j \leq L-1 \right) \\
-\ii\,( \mu - \ii \gamma)\,\hat{a}_{L} & \left( j=L \right)
\end{cases}.
\end{split} \end{equation}
Hence the Majorana edge modes given by Eq.~(\ref{eq: Majorana edge modes}) satisfy
\begin{equation} \begin{split}
[\hat{H}_{\rm PT},\,\hat{\Psi}_{\rm zero}^{\rm L}]
\propto - 2\ii J \left( 1 + \frac{\gamma^{2}}{\mu^{2}} \right) \left( - \frac{\mu}{2J} \right)^{L} \hat{b}_{L},~~
[\hat{H}_{\rm PT},\,\hat{\Psi}_{\rm zero}^{\rm R}]
\propto 2\ii J \left( 1 + \frac{\gamma^{2}}{\mu^{2}} \right) \left( - \frac{\mu}{2J} \right)^{L} \hat{a}_{1}
\end{split} \end{equation}
for arbitrary finite $L$ and the coefficients are exponentially small in $L$. We note that this exponential modification is the same as the Hermitian Majorana chain. In addition, the complex edge modes given by Eq.~(\ref{eq: complex edge modes}) satisfy
\begin{equation} \begin{split}
[\hat{H}_{\rm PT},\,\hat{\Psi}_{\rm complex}^{\rm L}] - E\,\hat{\Psi}_{\rm complex}^{\rm L}
\propto - \lambda^{L-1} \left[ \gamma y\,\hat{a}_{L} + \ii \left( \mu+\ii \gamma \right) \left( \frac{\ii \gamma}{\mu} + \frac{4J^{2}}{\gamma \left( 2\ii \mu - \gamma \right)} \right) \hat{b}_{L} \right]
\end{split} \end{equation}
for arbitrary finite $L$ and the coefficients are exponentially small in $L$.

\section{Pseudo-Hermiticity operator}
	\label{appendix: pseudo-Hermiticity}

We prove that $\hat{\eta}$ given by Eq.~(\ref{eq: pseudo-Hermiticity}) satisfies the pseudo-Hermiticity algebra given by Eq.~(\ref{eq: pseudo-Hermiticity - definition}). We here introduce $\alpha := \hop+\Delta,\,\beta := \hop-\Delta$. First we note the following relations:
\begin{small} \begin{equation} \begin{split}
&\left[ \hat{\eta},\,\hat{b}_{2n} \hat{a}_{2n+1} \right] \\ 
&= \left( 1 - \frac{\ii \gamma}{\alpha}\,\hat{a}_{1} \hat{a}_{2} \right) \cdots \left[ \left( 1 - \frac{\ii \gamma}{\alpha}\,\hat{b}_{2n} \hat{b}_{2n+1} \right) \left( 1 - \frac{\ii \gamma}{\alpha}\,\hat{a}_{2n+1} \hat{a}_{2n+2} \right),\,\hat{b}_{2n} \hat{a}_{2n+1} \right] \cdots \left( 1 - \frac{\ii \gamma}{\alpha}\,\hat{b}_{L-1} \hat{b}_{L} \right) \\
&= - \frac{2\ii \gamma}{\alpha} \left( 1 - \frac{\ii \gamma}{\alpha}\,\hat{a}_{1} \hat{a}_{2} \right) \cdots \left( 1 - \frac{\ii \gamma}{\alpha}\,\hat{a}_{2n-1} \hat{a}_{2n} \right) \left( \hat{a}_{2n+1} \hat{b}_{2n+1} - \hat{b}_{2n} \hat{a}_{2n+2} \right) \left( 1 - \frac{\ii \gamma}{\alpha}\,\hat{b}_{2n+2} \hat{b}_{2n+3} \right) \cdots \left( 1 - \frac{\ii \gamma}{\alpha}\,\hat{b}_{L-1} \hat{b}_{L} \right), \\
&\left[ \hat{\eta},\,\hat{b}_{2n+1} \hat{a}_{2n+2} \right] \\ 
&= \left( 1 - \frac{\ii \gamma}{\alpha}\,\hat{a}_{1} \hat{a}_{2} \right) \cdots \left[ \left( 1 - \frac{\ii \gamma}{\alpha}\,\hat{b}_{2n} \hat{b}_{2n+1} \right) \left( 1 - \frac{\ii \gamma}{\alpha}\,\hat{a}_{2n+1} \hat{a}_{2n+2} \right),\,\hat{b}_{2n+1} \hat{a}_{2n+2} \right] \cdots \left( 1 - \frac{\ii \gamma}{\alpha}\,\hat{b}_{L-1} \hat{b}_{L} \right) \\
&= + \frac{2\ii \gamma}{\alpha} \left( 1 - \frac{\ii \gamma}{\alpha}\,\hat{a}_{1} \hat{a}_{2} \right) \cdots \left( 1 - \frac{\ii \gamma}{\alpha}\,\hat{a}_{2n-1} \hat{a}_{2n} \right) \left( \hat{a}_{2n+1} \hat{b}_{2n+1} - \hat{b}_{2n} \hat{a}_{2n+2} \right) \left( 1 - \frac{\ii \gamma}{\alpha}\,\hat{b}_{2n+2} \hat{b}_{2n+3} \right) \cdots \left( 1 - \frac{\ii \gamma}{\alpha}\,\hat{b}_{L-1} \hat{b}_{L} \right),
\end{split} \end{equation} \end{small}
leading to $[ \hat{\eta}, ( \hat{b}_{2n} \hat{a}_{2n+1} + \hat{b}_{2n+1} \hat{a}_{2n+2} ) ] = 0$ for $n=1,2,\cdots,(L-3)/2$. In a similar manner, we obtain $[ \hat{\eta}, ( \hat{a}_{2n-1} \hat{b}_{2n} + \hat{a}_{2n} \hat{a}_{2n+1} ) ] = 0$ for $n=1,2,\cdots,(L-1)/2$. Next we note that
\begin{small} \begin{equation} \begin{split}
\left\{ \hat{\eta},\,\hat{a}_{1} \hat{b}_{1} \right\} 
&= \left\{ 1 - \frac{\ii \gamma}{\alpha}\,\hat{a}_{1} \hat{a}_{2},\,\hat{a}_{1} \hat{b}_{1} \right\} \left( 1 - \frac{\ii \gamma}{\alpha}\,\hat{b}_{2} \hat{b}_{3} \right) \cdots \left( 1 - \frac{\ii \gamma}{\alpha}\,\hat{b}_{L-1} \hat{b}_{L} \right)
=  \left( 2\,\hat{a}_{1} \hat{b}_{1} \right) \left( 1 - \frac{\ii \gamma}{\alpha}\,\hat{b}_{2} \hat{b}_{3} \right) \cdots \left( 1 - \frac{\ii \gamma}{\alpha}\,\hat{b}_{L-1} \hat{b}_{L} \right), \\
\left\{ \eta,\,\hat{a}_{L} \hat{b}_{L} \right\} 
&= \left( 1 - \frac{\ii \gamma}{\alpha}\,\hat{a}_{1} \hat{a}_{2} \right) \cdots \left( 1 - \frac{\ii \gamma}{\alpha}\,\hat{a}_{L-2} \hat{a}_{L-1} \right) \left\{ 1 - \frac{\ii \gamma}{\alpha}\,\hat{b}_{L-1} \hat{b}_{L},\,\hat{a}_{L} \hat{b}_{L} \right\}
= \left( 1 - \frac{\ii \gamma}{\alpha}\,\hat{a}_{1} \hat{a}_{2} \right) \cdots \left( 1 - \frac{\ii \gamma}{\alpha}\,\hat{a}_{L-2} \hat{a}_{L-1} \right) \left( 2\,\hat{a}_{L} \hat{b}_{L} \right).
\end{split} \end{equation} \end{small}
Hence we have
\begin{eqnarray}
\hat{\eta} \hat{H}_{\rm PT} - \hat{H}_{\rm PT}^{\dag} \hat{\eta}
&=& \frac{\ii \alpha}{2} \left[ \hat{\eta},\,\sum_{j=1}^{L-1} \hat{b}_{j} \hat{a}_{j+1} \right]
- \frac{\ii \beta}{2} \left[ \hat{\eta},\,\sum_{j=1}^{L-1} \hat{a}_{j} \hat{b}_{j+1} \right]
+ \frac{\gamma}{2} \left\{ \hat{\eta},\,\left( \hat{a}_{1} \hat{b}_{1} - \hat{a}_{L} \hat{b}_{L} \right) \right\} \nonumber \\
&=& \frac{\ii \alpha}{2} \left[ \hat{\eta}, \left( \hat{b}_{1} \hat{a}_{2} + \hat{b}_{L-1} \hat{a}_{L} \right) \right] 
+ \frac{\gamma}{2} \left\{ \hat{\eta},\,\left( \hat{a}_{1} \hat{b}_{1} - \hat{a}_{L} \hat{b}_{L} \right) \right\} = 0.
\end{eqnarray}

We here remark that
\begin{equation}
\hat{\eta}'
= \left( 1 - \frac{\ii \gamma}{\beta} \hat{b}_{1} \hat{b}_{2} \right) \left( 1 - \frac{\ii \gamma}{\beta} \hat{a}_{2} \hat{a}_{3} \right) \cdots \left( 1 - \frac{\ii \gamma}{\beta} \hat{b}_{L-2} \hat{b}_{L-1} \right) \left( 1 - \frac{\ii \gamma}{\beta} \hat{a}_{L-1} \hat{a}_{L} \right) 
\end{equation}
also satisfies $\hat{\eta}' \hat{H}_{\rm PT} - \hat{H}_{\rm PT}^{\dag} \hat{\eta}' = 0$, but $\hat{\eta}'$ ceases to be positive at $\gamma = \beta$, which is below the PT-transition point ($\gamma = \alpha > \beta$). 

In general, it is still difficult to have the accompanying Hermitian Hamiltonian given by Eq.~(\ref{eq: pseudo-Hermiticity - Hermitian}) for $\hat{H}_{\rm PT}$ with $L$ sites even if $\hat{\eta}$ is exactly determined. Nevertheless, $\hat{H}_{\rm H}$ can also be analytically obtained in our model since $\hat{\eta}$ in Eq.~(\ref{eq: pseudo-Hermiticity - definition}) is given by the product of $L-1$ commuting operators. We note that 
\begin{small} \begin{equation} \begin{split}
\left( 1 - \frac{\ii \gamma}{\alpha} \hat{a}_{j} \hat{a}_{j+1} \right)^{1/2}
&= \frac{1}{2} \left[ \left( \sqrt{1 + \frac{\gamma}{\alpha}} + \sqrt{1 - \frac{\gamma}{\alpha}} \right) -  \left( \sqrt{1 + \frac{\gamma}{\alpha}} - \sqrt{1 - \frac{\gamma}{\alpha}} \right) \ii \hat{a}_{j} \hat{a}_{j+1} \right] 
, \\
\left( 1 - \frac{\ii \gamma}{\alpha} \hat{a}_{j} \hat{a}_{j+1} \right)^{-1/2}
&= \frac{1}{2 \sqrt{1-\left( \gamma/\alpha \right)^{2}}} \left[ \left( \sqrt{1 + \frac{\gamma}{\alpha}} + \sqrt{1 - \frac{\gamma}{\alpha}} \right) + \left( \sqrt{1 + \frac{\gamma}{\alpha}} - \sqrt{1 - \frac{\gamma}{\alpha}} \right) \ii \hat{a}_{j} \hat{a}_{j+1} \right].
\end{split} \end{equation} \end{small}
We thus have 
\begin{equation} \begin{split}
\hat{\eta}^{1/2} \left( \hat{b}_{2n} \hat{a}_{2n+1} + \hat{b}_{2n+1} \hat{a}_{2n+2} \right) \hat{\eta}^{-1/2}
&= \hat{b}_{2n} \hat{a}_{2n+1} + \hat{b}_{2n+1} \hat{a}_{2n+2}~~~\left( n=1,2,\cdots, ( L-3 )/2 \right), \\
\hat{\eta}^{1/2} \left( \hat{a}_{2n-1} \hat{b}_{2n} + \hat{a}_{2n} \hat{b}_{2n+1} \right) \hat{\eta}^{-1/2}
&= \hat{a}_{2n-1} \hat{b}_{2n} + \hat{a}_{2n} \hat{b}_{2n+1}~~~\left( n=1,2,\cdots, ( L-1 )/2 \right),
\end{split} \end{equation}
and 
\begin{equation} \begin{split}
\hat{\eta}^{1/2} \left( \alpha\,\hat{b}_{1} \hat{a}_{2} - \ii \gamma\,\hat{a}_{1} \hat{b}_{1} \right) \hat{\eta}^{-1/2}
= \sqrt{\alpha^{2}-\gamma^{2}}\,\hat{b}_{1} \hat{a}_{2},~~
\hat{\eta}^{1/2} \left( \alpha\,\hat{b}_{L-1} \hat{a}_{L} + \ii \gamma\,\hat{a}_{L} \hat{b}_{L} \right) \hat{\eta}^{-1/2}
= \sqrt{\alpha^{2}-\gamma^{2}}\,\hat{b}_{L-1} \hat{a}_{L}.
\end{split} \end{equation}
Therefore $H_{\rm H}$ is obtained as Eq.~(\ref{eq: Hamiltonian - Hermitian}).

\section{Non-Hermitian free fermion numerics}
	\label{appendix: fermion numerics}

We consider the diagonalization of a general non-Hermitian and noninteracting (quadratic) fermionic system $\hat{H}$ with chiral symmetry, including the PT-symmetric Majorana chain $\hat{H}_{\rm PT}$. The Hamiltonian $\hat{H}$ can be expressed as
\begin{equation}
\hat{H}
= \left( \begin{array}{@{\,}cc@{\,}}
      \hat{\bm c}^{\dag} & \hat{\bm c}
    \end{array} \right) {\cal H}_{\rm BdG} \left( \begin{array}{@{\,}c@{\,}}
      \hat{\bm c} \\ \hat{\bm c}^{\dag}
    \end{array} \right),
\end{equation}
where $\hat{\bm c} := (\hat{c}_{1},\cdots,\hat{c}_{L})$ ($\hat{\bm c}^{\dag} := (\hat{c}_{1}^{\dag},\cdots,\hat{c}_{L}^{\dag})$) is a vector of annihilation (creation) operators. The $2L \times 2L$ matrix ${\cal H}_{\rm BdG}$ has chiral symmetry, i.e., $\left( \tau_{y} \otimes I_{L} \right) {\cal H}_{\rm BdG} \left( \tau_{y} \otimes I_{L} \right) = - {\cal H}_{\rm BdG}$ ($\tau_{y}$ is the Pauli matrix and $I_{L}$ is the $L \times L$ identity matrix), and its spectrum can be represented as $( -E_{1},\,\cdots,-E_{L},\,+E_{1},\,\cdots,+E_{L} )$ (${\rm Re}\,E_{j} > 0$). If the corresponding right eigenvectors are denoted by $V := (\ket{\varphi_{1}},\,\cdots,\ket{\varphi_{2L}})$, $V$ is nonunitary and $V^{-1} = (\bra{\chi_{1}},\,\cdots,\bra{\chi_{2L}})^{T}$, where $\bra{\chi_{j}}$ is the left eigenvector normalized by $\braket{\chi_{i} | \varphi_{j}} = \delta_{ij}$~\cite{Brody-14}. In addition, $V$ satisfies $\left( \tau_{y} \otimes I_{L} \right) V \left( \tau_{y} \otimes I_{L} \right) = V$ due to the presence of chiral symmetry and thus takes the form of
\begin{equation}
V = \left( \begin{array}{@{\,}cc@{\,}}
      A & -B \\ B & A
    \end{array} \right),
\end{equation}
where $A$ and $B$ are $L \times L$ matrices. We define the quasiparticles $\hat{\bm p}^{\dag} := (\hat{p}_{1}^{\dag},\cdots,\hat{p}_{L}^{\dag})$ and $\hat{\bm q} := (\hat{q}_{1},\cdots,\hat{q}_{L})$ by
\begin{equation}
\left( \begin{array}{@{\,}cc@{\,}}
      \hat{\bm q} & \hat{\bm p}^{\dag}
    \end{array} \right) := \left( \begin{array}{@{\,}cc@{\,}}
      \hat{\bm c}^{\dag} & \hat{\bm c}
    \end{array} \right) V,~~
\left( \begin{array}{@{\,}c@{\,}}
      \hat{\bm p}^{\dag} \\ \hat{\bm q}
    \end{array} \right) := V^{-1} \left( \begin{array}{@{\,}c@{\,}}
      \hat{\bm c} \\ \hat{\bm c}^{\dag}
    \end{array} \right),
\end{equation}
which satisfy the anticommutation relations 
\begin{equation}
\{ \hat{p}_{i}^{\dag},\,\hat{q}_{j} \} = \delta_{ij},~\{ \hat{p}_{i}^{\dag},\,\hat{p}_{j}^{\dag} \} = \{ \hat{q}_{i},\,\hat{q}_{j} \} = 0.
\end{equation}
Then $\hat{H}$ is diagonalized with $\hat{\bm p}^{\dag},\,\hat{\bm q}$ as
\begin{equation} \begin{split}
\hat{H}
= \left( \begin{array}{@{\,}cc@{\,}}
      \hat{\bm c}^{\dag} & \hat{\bm c}
    \end{array} \right) V \left( V^{-1} {\cal H}_{\rm BdG} V \right) V^{-1} \left( \begin{array}{@{\,}c@{\,}}
      \hat{\bm c} \\ \hat{\bm c}^{\dag}
    \end{array} \right)
= \left( \begin{array}{@{\,}cc@{\,}}
      \hat{\bm q} & \hat{\bm p}^{\dag}
    \end{array} \right) \left[ {\rm diag} \left( E_{j} \right) \right] \left( \begin{array}{@{\,}c@{\,}}
      \hat{\bm p}^{\dag} \\ \hat{\bm q}
    \end{array} \right)
= \sum_{j=1}^{L} \left( 2E_{j} \right) \hat{p}_{j}^{\dag} \hat{q}_{j} - \sum_{j=1}^{L} E_{j}.
\end{split} \end{equation}
If the vacuum for the quasiparticles $\hat{\bm q}$ is defined as $\ket{\Omega}$ (i.e., $\hat{q}_{j} \ket{\Omega} = 0$ for all $j$), we have
\begin{equation} \begin{split}
\hat{H}\,( \hat{p}_{j}^{\dag} \ket{\Omega} )
= \left( 2E_{j} - E_{0} \right) ( \hat{p}_{j}^{\dag} \ket{\Omega} ),~~
\hat{H}^{\dag}\,( \hat{q}_{j}^{\dag} \ket{\Omega} )
= \left( 2E_{j}^{*} - E_{0}^{*} \right) ( \hat{q}_{j}^{\dag} \ket{\Omega} ),
\end{split} \end{equation}
with $E_{0} := \sum_{j=1}^{L} E_{j}$. In other words, $\hat{\bm p}^{\dag}$ and $\hat{\bm q}^{\dag}$ create right and left single-particle eigenstates, respectively.

We next consider the dynamics of $\hat{H}$. We take an initial state as $\ket{\Psi \left( 0 \right)} = \sum_{j=1}^{L} \lambda_{j} \hat{p}_{j}^{\dag} \ket{\Omega}$ and then the unnormalized many-particle wavefunction $\ket{\tilde{\Psi} \left( t \right)} := e^{-\ii \hat{H} t} \ket{\Psi \left( 0 \right)}$ evolves into
\begin{equation} \begin{split}
\ket{\tilde{\Psi} \left( t \right)}
= e^{\ii E_{0} t} \sum_{j=1}^{L} \lambda_{j} \left( t \right) \hat{p}_{j}^{\dag} \ket{\Omega},
\end{split} \end{equation}
with $\lambda_{j} \left( t \right) := \lambda_{j} e^{-2\ii E_{j} t}$. We investigate the time evolution of the particle current between sites $j-1$ and $j$:
\begin{equation} \begin{split}
I_{j} \left( t \right)
:= \braket{\Psi \left( t \right) | \hat{I}_{j} | \Psi \left( t \right)}
= \frac{\braket{\tilde{\Psi} \left( t \right) | -\ii \hop\,(\hat{c}_{j}^{\dag} \hat{c}_{j-1} - \hat{c}_{j-1}^{\dag} \hat{c}_{j}) | \tilde{\Psi} \left( t \right)}}{\braket{\tilde{\Psi} \left( t \right) | \tilde{\Psi} \left( t \right)}}
= - 2\hop \times \frac{{\rm Im} \left[ \braket{\tilde{\Psi} \left( t \right) | \hat{c}_{j-1}^{\dag} \hat{c}_{j} | \tilde{\Psi} \left( t \right)} \right]}{\braket{\tilde{\Psi} \left( t \right) | \tilde{\Psi} \left( t \right)}}.
\end{split} \end{equation}
We note that
\begin{equation} \begin{split}
\braket{\tilde{\Psi} \left( t \right) | \tilde{\Psi} \left( t \right)}
= e^{-2\,{\rm Im}\left[ E_{0} \right] t} \sum_{m,n=1}^{L} \lambda_{m}^{*} \left( t \right) \braket{\Omega | \hat{p}_{m} \hat{p}^{\dag}_{n} | \Omega} \lambda_{n} \left( t \right)
= e^{-2\,{\rm Im}\left[ E_{0} \right] t} \sum_{m,n=1}^{L} \lambda_{m}^{*} \left( t \right) X_{mn} \lambda_{n} \left( t \right),
\end{split} \end{equation}
where $\hat{p}_{m}$ is expanded as $\hat{p}_{m} = \sum_{l=1}^{L}\,[ X_{ml} \hat{q}_{l} + Y_{ml} \hat{q}_{l}^{\dag} ]$. The coefficient matrix $X_{mn}$ is determined from
\begin{equation} \begin{split}
\delta_{mn}
= \{ \hat{p}_{m},\,\hat{q}_{n}^{\dag} \}
= \sum_{l=1}^{L} X_{ml} \{ \hat{q}_{l},\,\hat{q}_{n}^{\dag} \}
= \sum_{l=1}^{L} X_{ml} \left( A^{T} A^{*} + B^{T} B^{*} \right)_{ln}
\end{split} \end{equation}
to be $X_{mn} = [ (A^{T}A^{*} + B^{T}B^{*})^{-1} ]_{mn}$. Here the appearance of off-diagonal elements in $X_{mn}$ is a consequence of the nonorthogonality in non-Hermitian systems; indeed, we have $X_{mn} = \delta_{mn}$ in the Hermitian limit. Next we notice that  
\begin{equation}
\braket{\tilde{\Psi} \left( t \right) | \hat{c}_{j-1}^{\dag} \hat{c}_{j} | \tilde{\Psi} \left( t \right)}
= e^{-2\,{\rm Im}\left[ E_{0} \right] t} \sum_{m,n=1}^{L} \lambda_{m}^{*} \left( t \right) M_{mn}^{j} \lambda_{n} \left( t \right),
\end{equation}
with 
\begin{equation} \begin{split}
M_{mn}^{j} 
&:= \braket{\Omega | \hat{p}_{m} \hat{c}_{j-1}^{\dag} \hat{c}_{j} \hat{p}^{\dag}_{n} | \Omega} \\
&= \sum_{k,l=1}^{L} \braket{\Omega | \hat{p}_{m} \left( A_{j-1,k}^{*} \hat{p}_{k} - B_{j-1,k}^{*} \hat{q}_{k}^{\dag} \right) \left( A_{jl} \hat{p}_{l}^{\dag} - B_{jl} \hat{q}_{l} \right) \hat{p}^{\dag}_{n} | \Omega} \\
&= \sum_{k,l=1}^{L} A_{j-1,k}^{*} A_{jl} \braket{\Omega | \hat{p}_{m} \hat{p}_{k} \hat{p}_{l}^{\dag} \hat{p}_{n}^{\dag} |\Omega}
- B_{j-1,m}^{*} \sum_{l=1}^{L} A_{jl} \braket{\Omega | \hat{p}_{l}^{\dag} \hat{p}_{n}^{\dag} |\Omega}
- B_{jn} \sum_{k=1}^{L} A_{j-1,k}^{*} \braket{\Omega | \hat{p}_{m} \hat{p}_{k} |\Omega}
+ B_{j-1,m}^{*} B_{jn}.
\end{split} \end{equation}
Here,
\begin{equation} \begin{split}
Z_{ij} 
:= \{ \hat{q}_{i},\,\hat{p}_{j} \}
= \sum_{m,n=1}^{L} \{ A_{mi} \hat{c}_{m}^{\dag} + B_{mi} \hat{c}_{m},\,-B_{nj}^{*} \hat{c}_{n} + A_{nj}^{*} \hat{c}_{n}^{\dag} \}
= \left( B^{T} A^{*} - A^{T} B^{*} \right)_{ij},
\end{split} \end{equation}
and we have
\begin{equation} \begin{split}
\braket{\Omega | \hat{p}_{m} \hat{p}_{k} |\Omega}
= \sum_{l=1}^{L} X_{ml} \braket{\Omega | \hat{q}_{l} \hat{p}_{k} |\Omega}
= \sum_{l=1}^{L} X_{ml} Z_{lk}
= \left( X Z \right)_{mk}.
\end{split} \end{equation}
Moreover, using Wick's theorem, 
\begin{equation} \begin{split}
\braket{\Omega | \hat{p}_{m} \hat{p}_{k} \hat{p}_{l}^{\dag} \hat{p}_{n}^{\dag} |\Omega}
&= \braket{\Omega | \hat{p}_{m} \hat{p}_{k} | \Omega} \braket{\Omega | \hat{p}_{l}^{\dag} \hat{p}_{n}^{\dag} | \Omega}
- \braket{\Omega | \hat{p}_{m} \hat{p}_{l}^{\dag} | \Omega} \braket{\Omega | \hat{p}_{k} \hat{p}_{n}^{\dag} | \Omega}
+ \braket{\Omega | \hat{p}_{m} \hat{p}_{n}^{\dag} | \Omega} \braket{\Omega | \hat{p}_{k} \hat{p}_{l}^{\dag} | \Omega} \\
&= \left( X Z \right)_{mk} ( (X Z)^{\dag} )_{ln}
- X_{ml} X_{kn} + X_{mn} X_{kl}.
\end{split} \end{equation}
Hence $M_{mn}^{j}$ is determined as 
\begin{equation} \begin{split}
M_{mn}^{j}
&= [ A^{*} (X Z)^{T} ]_{j-1,m}\,[ A\,(X Z)^{\dag} ]_{jn}
- [ A X^{T} ]_{jm} [ A^{*} X ]_{j-1,n}
+ [ A^{*} X A^{T} ]_{j-1,j} X_{mn} \\
&~~~ - B_{j-1,m}^{*} ( A\,(X Z)^{\dag} )_{jn}
- ( A^{*} (X Z)^{T} )_{j-1,m} B_{jn}
+ B_{j-1,m}^{*} B_{jn},
\end{split} \end{equation}
and the particle current is obtained as
\begin{equation}
I_{j} \left( t \right)
:= \braket{\Psi \left( t \right) | \hat{I}_{j} | \Psi \left( t \right)}
= -2\hop \times \frac{\sum_{m,n=1}^{L} {\rm Im} \left[ \lambda_{m}^{*} \left( t \right) M_{mn}^{j} \lambda_{n} \left( t \right) \right]}{\sum_{m,n=1}^{L} \lambda_{m}^{*} \left( t \right) X_{mn} \lambda_{n} \left( t \right)}.
\end{equation}

\section{Details on PT-symmetric fermionic system with two sites}
	\label{appendix: two sites}

Equation (\ref{eq: Hamiltonian - two sites}) can be represented as 
\begin{equation}
\hat{H}_{\rm PT}^{(2)}
:= - \left( \begin{array}{@{\,}cc@{\,}}
      \hat{c}_{1}^{\dag} & \hat{c}_{2}^{\dag}
    \end{array} \right) {\cal H} \left( \begin{array}{@{\,}c@{\,}}
      \hat{c}_{1} \\ \hat{c}_{2}
    \end{array} \right),~~
{\cal H} := \left( \begin{array}{@{\,}cc@{\,}}
      \ii \gamma & \hop \\ \hop & -\ii \gamma
    \end{array} \right).
\end{equation}
The eigenvalues of ${\cal H}$ are $E = \pm \sqrt{\hop^{2} - \gamma^{2}}$, and thus the PT-transition point lies at $\gamma = \hop$. We first consider the PT-unbroken phase ($\gamma < \hop$), where the normalized eigenvectors are given by
\begin{equation}
\ket{\pm}
= \frac{1}{\sqrt{2}} \left( \begin{array}{@{\,}c@{\,}}
      1 \\ \pm \sqrt{1 - \left( \gamma/\hop \right)^{2}} - \ii \gamma/\hop
    \end{array} \right).
\end{equation}
We define the quasiparticles by
\begin{equation}
\left( \begin{array}{@{\,}cc@{\,}}
      \hat{p}_{1}^{\dag} & \hat{p}_{2}^{\dag}
    \end{array} \right) := \left( \begin{array}{@{\,}cc@{\,}}
      \hat{c}_{1}^{\dag} & \hat{c}_{2}^{\dag}
    \end{array} \right) V,~~
\left( \begin{array}{@{\,}c@{\,}}
      \hat{q}_{1} \\ \hat{q}_{2}
    \end{array} \right) := V^{-1} \left( \begin{array}{@{\,}c@{\,}}
      \hat{c}_{1} \\ \hat{c}_{2}
    \end{array} \right),
    	\label{eq: p&q definition}
\end{equation}
with $V := \left( \ket{+},\,\ket{-} \right)$. The operators which characterize these quasiparticles satisfy the anticommutation relations given by Eq.~(\ref{eq: anticommutation - two sites}), and the Hamiltonian is diagonalized as Eq.~(\ref{eq: Hamiltonian diagonal - two-sites}). The relationship between $\hat{p}_{j}$ and $\hat{q}_{j}$, which is given by Eq.~(\ref{eq: pq - unbroken - two-sites}), can be straightforwardly calculated from Eq.~(\ref{eq: p&q definition}). The unnormalized wavefunction $\ket{\tilde{\Psi} \left( t \right)} := e^{-\ii \hat{H}_{\rm PT} t} \ket{\Psi \left( 0 \right)}$ evolves into
\begin{equation}
\ket{\tilde{\Psi} \left( t \right)}
= \left( \lambda_{1} e^{\ii \sqrt{\hop^{2}-\gamma^{2}} t} \hat{p}_{1}^{\dag} + \lambda_{2} e^{-\ii \sqrt{\hop^{2}-\gamma^{2}} t} \hat{p}_{2}^{\dag} \right) \ket{\Omega},
\end{equation}
and we obtain
\begin{equation} \begin{split}
\braket{\tilde{\Psi} \left( t \right) | \tilde{\Psi} \left( t \right)}
&= 1 + \frac{2\gamma}{\hop} {\rm Im} \left[ g \lambda_{1}^{*} \lambda_{2} e^{-2\ii \sqrt{\hop^{2}-\gamma^{2}}t} \right], \\
-\ii \braket{\tilde{\Psi} \left( t \right) | ( \hat{c}_{2}^{\dag} \hat{c}_{1} - \hat{c}_{1}^{\dag} \hat{c}_{2} ) | \tilde{\Psi} \left( t \right)}
&= {\rm Im} \left[ g \right] + 2\,{\rm Im} \left[ g \lambda_{1}^{*} \lambda_{2} e^{-2\ii \sqrt{\hop^{2} - \gamma^{2}}t} \right],
\end{split} \end{equation}
which leads to Eq.~(\ref{current - unbroken - two-sites}). In the PT-broken phase ($\gamma > \hop$), on the other hand, the normalized eigenvectors of ${\cal H}$ are
\begin{equation} \begin{split}
\ket{\pm} = \frac{1}{\sqrt{N_{\pm}}} \left( \begin{array}{@{\,}c@{\,}}
      1 \\ \ii \left( \pm \sqrt{\left( \gamma/\hop \right)^{2}-1} - \gamma/\hop \right)
    \end{array} \right),~~
    N_{\pm} := \frac{2\gamma}{\hop} \left( \frac{\gamma}{\hop} \mp \sqrt{\left( \frac{\gamma}{\hop} \right)^{2} - 1} \right).
\end{split} \end{equation}
The unnormalized wavefunction evolves into 
\begin{equation}
\ket{\tilde{\Psi} \left( t \right)} = \left( \lambda_{1} e^{-\sqrt{\gamma^{2}-\hop^{2}} t} \hat{p}_{1}^{\dag} + \lambda_{2} e^{\sqrt{\gamma^{2}-\hop^{2}} t} \hat{p}_{2}^{\dag} \right) \ket{\Omega},
\end{equation}
and we obtain
\begin{equation} \begin{split}
\braket{\tilde{\Psi} \left( t \right) | \tilde{\Psi} \left( t \right)}
&= \left| \lambda_{1} \right|^{2} e^{-2 \sqrt{\gamma^{2}-\hop^{2}} t} + \left| \lambda_{2} \right|^{2} e^{2 \sqrt{\gamma^{2}-\hop^{2}t}} + \frac{2\hop}{\gamma}\,{\rm Re} \left[ \lambda_{1}^{*} \lambda_{2} \right], \\
-\ii \braket{\tilde{\Psi} \left( t \right) | ( \hat{c}_{2}^{\dag} \hat{c}_{1} - \hat{c}_{1}^{\dag} \hat{c}_{2} ) | \tilde{\Psi} \left( t \right)}
&= \frac{\hop}{\gamma} \left( \left| \lambda_{1} \right|^{2} e^{-2 \sqrt{\gamma^{2}-\hop^{2}}t} + \left| \lambda_{2} \right|^{2} e^{2 \sqrt{\gamma^{2}-\hop^{2}}t}
+ \frac{2 \gamma}{\hop}\,{\rm Re} \left[ \lambda_{1}^{*} \lambda_{2} \right] \right),
\end{split} \end{equation}
which leads to
\begin{equation} \begin{split}
\frac{\braket{\Psi \left( t \right) | \hat{I} | \Psi \left( t \right)}}{\hop}
&= \frac{\hop}{\gamma} \times \frac{\left| \lambda_{1} \right|^{2} e^{-2 \sqrt{\gamma^{2}-\hop^{2}}t} + \left| \lambda_{2} \right|^{2} e^{2 \sqrt{\gamma^{2}-\hop^{2}}t}
+ 2 \left( \gamma/\hop \right) {\rm Re} \left[ \lambda_{1}^{*} \lambda_{2} \right]}{\left| \lambda_{1} \right|^{2} e^{-2 \sqrt{\gamma^{2}-\hop^{2}} t} + \left| \lambda_{2} \right|^{2} e^{2 \sqrt{\gamma^{2}-\hop^{2}t}} + 2 \left( \hop/\gamma \right) {\rm Re} \left[ \lambda_{1}^{*} \lambda_{2} \right]}~~\left( t \to \infty \right).
\end{split} \end{equation}
Hence we obtain Eq.~(\ref{current - broken - two-sites}).
\twocolumngrid



\begin{thebibliography}{}
\bibitem{remark-complex-edge}
We have numerically confirmed that no modes with complex eigenenergies exist except for the complex edge modes $\hat{\Psi}_{\rm complex}$.
\bibitem{remark-pseudo-Hermiticity}
Although there are perturbative constructions~\cite{Bender-04, Korff-07, Fring-09}, they work only for small non-Hermiticity and break down near the PT-transition point.
\end{thebibliography}
\end{document}